\documentclass[manusxript]{aastex62}
\usepackage{CJK}

\received{Submitted to Astronomical Journal}
\revised{}
\accepted{}
\shorttitle{High Perihelion Objects}
\shortauthors{Li et al.}
\begin{document}
\begin{CJK*}{UTF8}{gbsn}

\title{Hubble Space Telescope Search for Activity in High Perihelion Objects}

\correspondingauthor{Jing Li}
\email{jli@igpp.ucla.edu}

\author{Jing Li (李京)}
\affil{Department of Earth, Planetary and Space Sciences,
UCLA,  595 Charles Young Drive East, Los Angeles, CA 90095-1567}

\author{David Jewitt}
\affiliation{Department of Earth, Planetary and Space Sciences,
UCLA,  595 Charles Young Drive East, Los Angeles, CA 90095-1567}
\affiliation{Department of Physics and Astronomy, UCLA, 
430 Portola Plaza, Box 951547,
Los Angeles, CA 90095-1547}

\author{Max Mutchler}
\affiliation{Space Telescope Science Institute, 3700 San Martin Drive, Baltimore, MD 21218}

\author{Jessica Agarwal}
\affiliation{Max Planck Institute for Solar System Research, Justus-von-Liebig-Weg 3, 37077 Göttingen, Germany}

\author{Harold Weaver}
\affiliation{The Johns Hopkins University Applied Physics Laboratory, 11100 Johns Hopkins Road, Laurel, Maryland 20723 }

\begin{abstract}

Solar system objects with perihelia beyond the orbit of Jupiter ($q >$ 5 AU) are too cold for  water ice  to generate an appreciable coma via sublimation.  Despite this, numerous high perihelion objects (HPOs) including many comets and recently escaped Kuiper belt objects (``Centaurs'') are observed to be active out at least to the orbit of Saturn ($q \sim$ 10 AU).  Peak  equilibrium temperatures at 10 AU ($\sim$125 K), while far  too low to sublimate water ice, are sufficient to sublimate super-volatiles such as CO and CO$_2$ ice.    Temperatures at 10 AU are also high enough to trigger the rapid crystallization of exposed amorphous ice, thus constituting another possible driver of distant activity.  While supervolatile ices can sublimate strongly (as $r_H^{-2}$) to at least Kuiper belt (30 AU) distances,  crystallization is an exponential function of temperature that cannot be sustained much beyond $\sim$10 AU.  The heliocentric dependence of the activity thus suggests an observational test.  If activity in high perihelion objects is triggered by crystallization, then no examples of activity should be found with perihelia $q >>$ 10 AU.  If, on the other hand, activity is due to free sublimation of exposed supervolatile ices, or another cause, then distant activity might be detected.  We obtained  sensitive, high resolution Hubble Space Telescope observations of HPOs to search for activity beyond the crystallization zone.  No examples of activity were detected in 53 objects with $q >$ 15 AU, consistent with the crystallization trigger hypothesis.  However, sensitivity limits are such that we cannot reject the alternative hypothesis that mass loss is driven by the sublimation of supervolatile ices.  We also searched for binary companions in our sample, finding none and setting an empirical 3$\sigma$ limit to the binary fraction of $<8$\%.  

 %We also use the data to conduct a search for binary Centaurs, setting limits to the properties of such objects and enabling a comparison with the binaries in the Kuiper belt source population.  
\end{abstract}



\section{Introduction} \label{sec:intro}

The Kuiper belt is a vast solar system repository located beyond Neptune, containing billions of ice-rich objects larger than 1 km and having a combined mass $\lesssim$ 0.1 $M_{\oplus}$ (Earth mass $M_{\oplus} = 6.0\times10^{24}$ kg). The belt is a relic of the formation epoch, and a reservoir holding some of the least thermally processed material in the solar system.  Some Kuiper belt objects are dynamically unstable on billion-year timescales, causing them  to drift into planet-approaching orbits and hence to be scattered across the solar system.  In particular,  the  Neptune-approaching ``scattered disk'' component of the Kuiper belt (prototype object 1996 TL66; \citet{1997Natur.387..573L}) is the probable source region from which most objects escape \citep{2008ApJ...687..714V}. Some escaped Kuiper belt objects  diffuse into orbits interior to that of Jupiter (semimajor axis $a_J =$ 5.203 AU,  eccentricity, $e_J$ = 0.048 and perihelion and aphelion distances of 4.953 AU and 5.453 AU, respectively) where the  equilibrium radiation temperatures are sufficiently high  that the dominant cometary volatile (water ice, with latent heat $L \sim 2\times10^6$ J kg$^{-1}$), begins to sublimate.  The resulting comae lead to a re-labelling of the objects as Jupiter family comets (JFCs).     \\

Objects evolving between the Kuiper belt and  JFC populations are collectively known as ``Centaurs'' (prototype object (2066) Chiron, Kowal et al.~1979).  Their most important dynamical characteristic is that Centaurs experience strong and frequent gravitational deflections by the major planets, causing them to be dynamically short-lived. The mean half life is $\sim$3 Myr according to \citet{2004MNRAS.355..321H}, and $\sim$7 Myr according to \citet{2003AJ....126.3122T}, but with a wide range from 1 Myr to 100 Myr. No single definition of the Centaurs exists.  We favor the practical  definition  that Centaurs are objects that have semimajor axes $a < a_N$ and perihelia, $q$, in the range $a_J < q < a_N$, where $a_J = $ 5.2 AU and $a_N$ = 30 AU are the semimajor axes of Jupiter's and Neptune's orbits, provided they are not trapped in mean-motion resonances with the giant planets \citep{2009AJ....137.4296J}.   Others (e.g.~Horner et al.~2004a) relax the semimajor axis limit, noting that the dynamical character is essentially unchanged whether the object crosses Neptune's orbit or not.  By these relaxed standards various Kuiper belt objects, specifically the Neptune-crossers in the scattered disk population, are included. \\

2060 Chiron (Kowal et al.~1979) is widely regarded as the prototype Centaur (strangely ignoring  29P/Schwassmann-Wachmann 1, which also has a perihelion outside Jupiter's orbit, a semimajor axis smaller than Neptune's, and which was discovered four decades before Chiron).  29P/Schwassmann-Wachmann 1 is continuously active, driven by the sublimation of carbon monoxide at prodigious rates (Senay and Jewitt 1994, Womack et al.~2017).  Chiron displays a  (transient) dust coma \citep{1990Icar...83....1H,1990AJ....100..913L,1990AJ....100.1323M} and evidence of CN gas emission \citep{2001Icar..150...94B}. Comet-like activity has subsequently proved to be a feature of many members of the Centaur population \citep{2009AJ....137.4296J}.  Activity beyond Jupiter suggests the sublimation of more volatile (lower latent heat) ices, but  other physical processes may play a role.  Abundant  ices present in comets include CO, CO$_2$, and less abundant N$_2$, all of which are volatile at temperatures found in the giant planet region of the solar system and which are, presumably, present in Centaurs.  Indeed, direct detection of CO gas has been made in Centaur 29P/Schwassmann-Wachmann 1 ($q$ = 5.7 AU) \citep{1994Natur.371..229S,2001Icar..150..140F,2008A&A...484..537G} and weak CO emission has also been reported  in 174P/Echeclus ($q$ = 5.8 AU) \citep{2017AJ....153..230W} as well as Chiron ($q$ = 8.5 AU) \citep{1999SoSyR..33..187W}.  On the other hand,  CO has not been detected in most of the Centaurs in which it was sought \citep{2001A&A...377..343B,2008AJ....135..400J,2017MNRAS.468.2897D}.  \\

%Two observations are puzzling in the context of  the sublimation of exposed supervolatiles.  They both point to an unexpected change in  Centaur properties occurring at perihelion distance $q \sim$ 10 AU.  First, t

A clue about the nature of distant activity is provided by the observation that the known active Centaurs have perihelia $q \lesssim$ 10 AU, corresponding roughly to the orbit of Saturn, while Centaurs with larger perihelia appear inactive  \citep{2009AJ....137.4296J}.  This radial segregation is intriguing because supervolatile sublimation should proceed strongly out to much larger distances ($>30$ AU, in case of CO). The free sublimation of CO and other supervolatile ices offers no reason to expect the cessation of cometary activity at distances beyond $\sim$10 AU.\\

%Activity is sometimes observed at larger heliocentric distances, $r_H$, but always in Centaurs with  perihelia $q \lesssim$ 10 AU, for which delayed propagation of heat acquired at perihelion provides a plausible explanation.   \\

%Second, the distribution of optical colors of  Centaurs  with $q \lesssim$ 10 AU is different from that of Centaurs with $q > $ 10 AU.  The former is  unimodal (color index B-R = 1.30$\pm$0.05) and consistent with the colors of their progeny, the short-period comets \citep{2015AJ....150..201J} . In contrast, Centaurs with larger perihelia exhibit a bimodal color distribution (including an additional ultrared component with B-R = 1.47$\pm$0.06) that is consistent with the colors of the precursor population in the  Kuiper belt scattered disk  \citep{2009AJ....137.4296J,2015AJ....150..201J}.   The coincidence between the critical perihelion distance for the onset of activity and the depletion of red-surfaced Centaurs suggests that one may be a consequence of the other.  For example, it has been noted that cometary activity can bury exposed ultrared surface material in a very short time \citep{2002AJ....123.1039J}.  \\

The crystallization of amorphous water ice offers a more natural, but unproved, explanation for the activation of distant objects near 10 AU.  The timescale for crystallization is an extremely strong function of temperature, and hence of heliocentric distance, given by

\begin{equation}
\tau_{cr}(T) = 3.0\times10^{-21} \exp \left[\frac{E_A}{kT}\right]
\label{crystal}
\end{equation}

\noindent where $E_A$ is the activation energy, $k$ is Boltzmann's Constant and $E_A$/$k$ = 5370 K \citep{1989pmcm.rept...65S}.  
For example, at the subsolar point on an object at  10 AU the crystallization timescale (decades) is comparable to the orbit period while at 30 AU it is  longer than the age of the solar system \citep{2009AJ....137.4296J,2012AJ....144...97G}.   Crystallization, in addition to being exothermic,  triggers the expulsion of trapped molecules,  including supervolatile species \citep{2007Icar..190..655B}.    The mass loss in this hypothesis is still driven by expansion of the suddenly freed supervolatiles, but their rate of release  is, somewhat counter-intuitively, controlled by the thermodynamics of water ice.  For this reason, the  distance and temperature dependence of activity driven by crystallization is distinct from that of freely sublimating supervolatile ices. \\

\begin{figure}
\epsscale{1.0}
\plotone{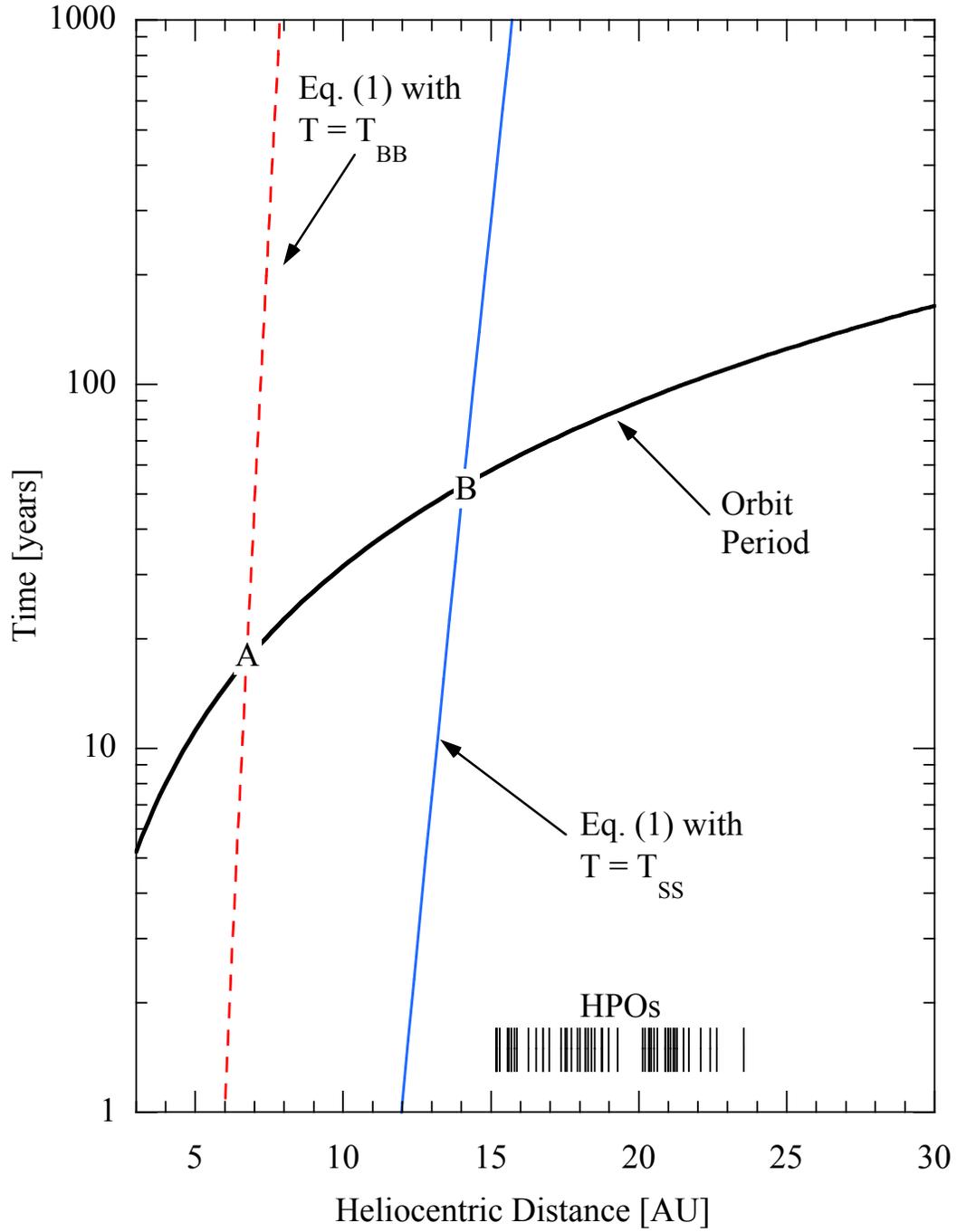}
\caption{Crystallization timescale (Equation \ref{crystal}) as a function of heliocentric distance  and two limiting temperature models.  The dashed red curve corresponds to the spherical blackbody temperature (i.e.~the low temperature limit) while the solid blue curve corresponds to the sub-solar temperature on a non-rotating nucleus.  The solid black curve shows the Keplerian orbit period.  Points $A$ and $B$ mark the inner and outer bounds of the region in which crystallization is expected.  The ``bar code'' at the bottom shows the perihelion distances of the 53 HPOs in our sample; all are more distant from the Sun than point B.  \label{tcrystal}}
\end{figure}

Figure \ref{tcrystal} shows solutions to Equation (\ref{crystal}) for two limiting cases of the surface temperature on a uniform, spherical body.  The  lowest possible radiative temperature  corresponds to the isothermal case, when power from the Sun is absorbed on one hemisphere but radiated uniformly from the entire surface.  The corresponding isothermal blackbody temperature is $T_{BB} = 278 r_H^{-1/2}$.  The highest local temperature is found at the subsolar point on a non-rotating body, and can be shown to equal $T_{SS} = 2^{1/2} T_{BB}$.  Low and high temperature solutions to Equation (\ref{crystal}) are shown in the figure as dashed red and solid blue lines, respectively.  We also show the Keplerian orbit period vs.~$r_H$ as a thick black line.  Points A and B mark the intersections of the curves and set the outer limits of the zones in which crystallization can occur on timescales comparable to the orbit period, for the low and high temperature cases.  The figure shows that crystallization can occur out to 14 AU, an inference confirmed by more sophisticated numerical calculations \citep{2012AJ....144...97G}.

The crystallization hypothesis is  attractive because laboratory experiments  (e.g. \citet{2009JPCA..11311174Z}) clearly show that the amorphous state is the natural state for ice accreted at low temperatures and pressures, as existed in the Sun's protoplanetary disk.  While physically plausible and widely assumed \citep{1997ApJ...478L.107P}, evidence for the existence of amorphous ice in solar system bodies is limited.   Proof of the abundance of amorphous ice would be scientifically valuable because, if the Centaurs are amorphous, then so must be their precursors in the Kuiper belt.  The amorphous/crystalline state of Kuiper belt ice in turn affects our understanding of the formation, composition and thermal evolution of these objects.\\

These considerations motivate the present study, in which we use the high resolution imaging capabilities of the Hubble Space Telescope (HST) to undertake a  sensitive search for activity in high-perihelion objects.  We reason that the detection of activity in objects with perihelia far beyond the  crystallization distance would invalidate the  hypothesis, and thus offer a scientifically useful, observational test. \\

\section{OBSERVATIONS} \label{sec:obs}
We employed  the 2.4 m diameter HST  for its superb angular resolution and sensitivity to near-nucleus dust.  Our observations were taken under  the Cycle 25  ``SNAP'' proposal GO 15344 using the Wide Field Camera 3 (WFC3).   WFC3 employs two 2000$\times$4000 pixel charge-coupled devices (CCDs) with an image scale 0.04\arcsec~pixel$^{-1}$, giving a field of view 162\arcsec$\times$162\arcsec~and a Nyquist sampled resolution of 0.08\arcsec.  We read out only half of one of the CCDs in order to optimize the observing efficiency and we used the  F350LP filter because it is broad (Full Width at Half Maximum (FWHM) $\sim$ 4758~\AA) and offers a high throughput needed to identify faint coma.  The effective central wavelength of F350LP, when used to observe a Sun-like (G2V) spectrum, is $\lambda_c$ = 6230~\AA.  For each target, we obtained two, consecutive images of 300 s duration with the telescope  tracked to follow the motion of the target. \\

\subsection{The Sample}

Our sample includes a broad mixture of  classical Centaurs ($a_J < q < a_N$ and $a < a_N$) and Neptune-crossing Kuiper belt objects ($a_J < q < a_N$, any $a$).  We refer to these collectively as high-perihelion objects (``HPOs'').  All HPOs were required to have $15 < q < $ 30 AU, comfortably larger than the $\sim$10 AU distance out to which activity has so far been detected, and to have ephemeris uncertainties $\le$10\arcsec~(obtained from the JPL ``Horizons'' web site\footnote{\url{https://ssd.jpl.nasa.gov/horizons.cgi}}) at the time of observation.  A total of 53 HPOs were successfully observed. Their orbital elements  are listed in Table (\ref{tobj}) in ascending order of the perihelion distance, $q$.   Figure \ref{tcrystal} shows the perihelia as short black lines, all clearly more distant from the Sun than the high-temperature critical point B.  \\

\begin{figure}
\epsscale{1.0}
\plotone{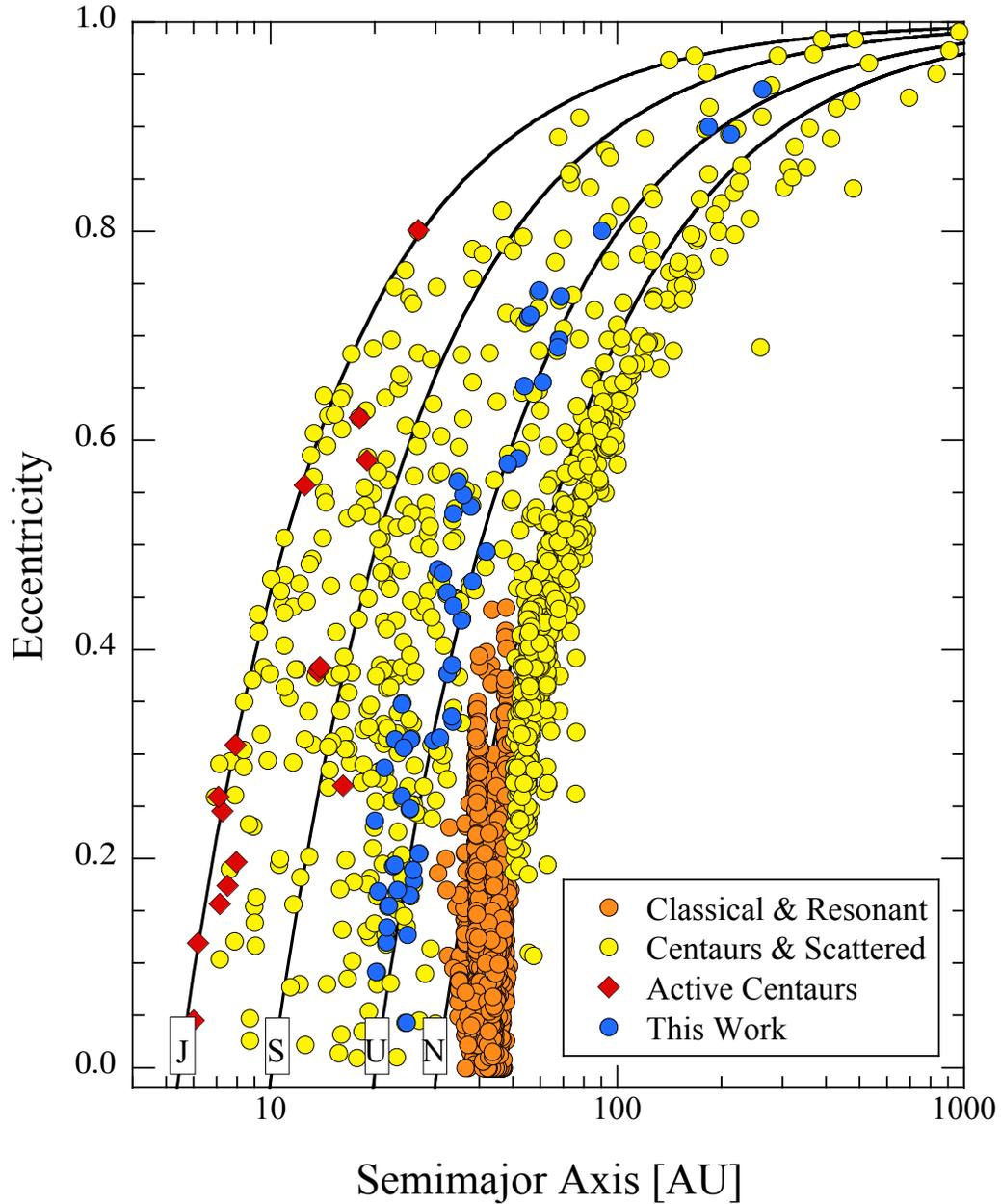}
\caption{Orbital eccentricity as a function of the semi-major axis. Colors represent different classes of objects. Centaurs observed in this SNAP/HST program are plotted as filled  blue circles. The classical and resonant Kuiper belt objects are in orange circles and Centaurs including scattered Kuiper Belt objects  are in yellow circles. They are found in the Minor Planet Center web site: https://minorplanetcenter.net/iau/mpc.html. Active Centaurs (red diamonds) are from \citet{2009AJ....137.4296J, 2012AJ....144...97G}. Solid curves show the loci of orbits having aphelion distances at Jupiter ($Q$ = 5.46AU), Saturn ($Q$ = 10.12 AU), Uranus ($Q$ = 20.11 AU) and Neptune ($Q$ = 30.33 AU). \label{ea}}
\end{figure}

\begin{figure}
\epsscale{1.0}
%\plotone{all_objs_ai.pdf}
\plotone{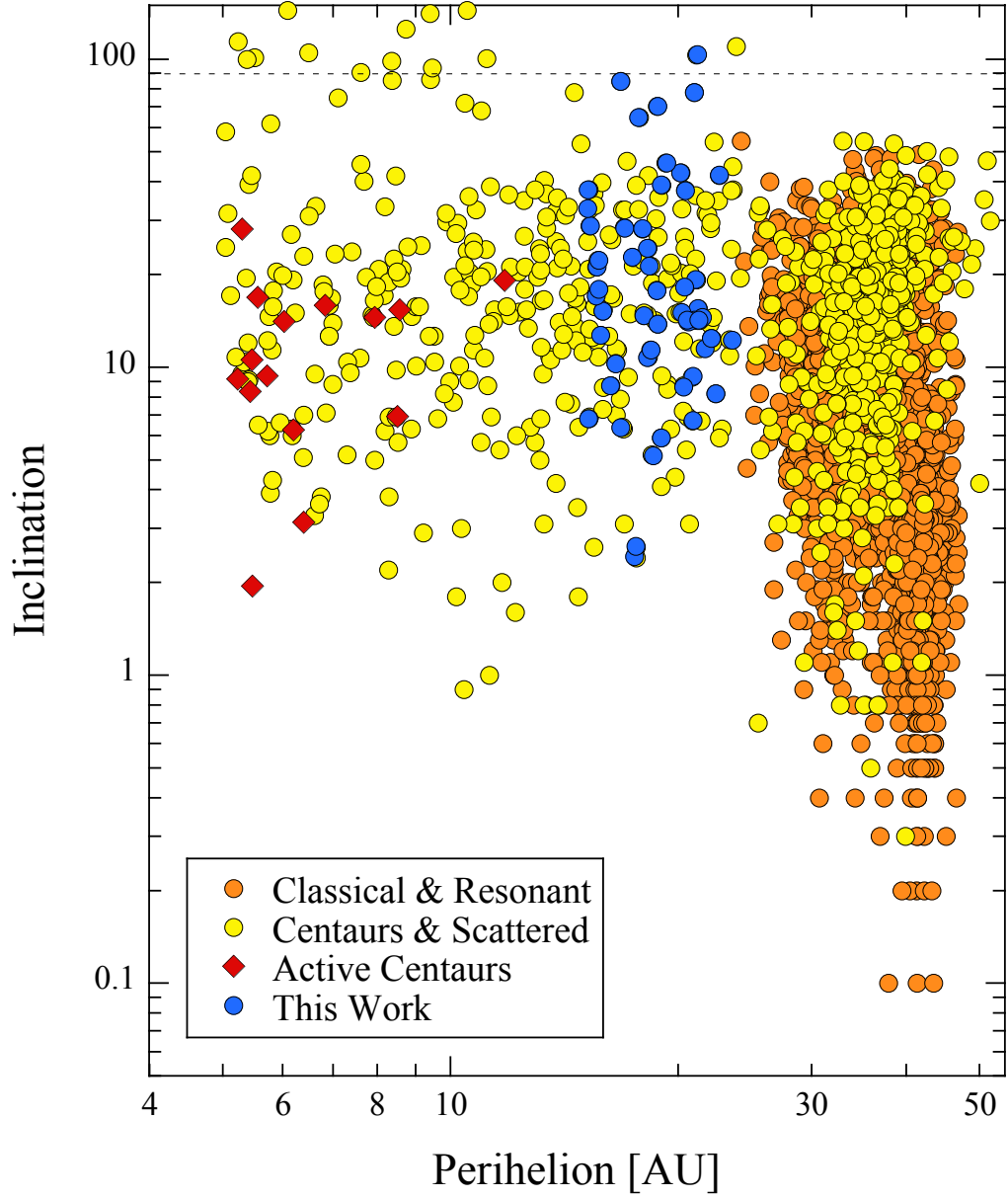}
\caption{Orbital inclination as a function of the perihelia. Different object classes are color-coded as in Figure (\ref{ea}). Objects above the dashed, horizontal line are retrograde (inclination $i >$  90\degr). Concentration of active objects with $q \lesssim$ 10 AU is evident.  \label{qi}}
\end{figure}

Figure \ref{ea} shows the object distribution in the semi-major axis versus eccentricity plane, with Trans-Neptunian objects (TNOs, combining the classical and resonant population)  plotted as orange circles\footnote{\url{https://www.minorplanetcenter.net/iau/lists/TNOs.html}}, and Centaurs and scattered objects as yellow circles\footnote{\url{https://www.minorplanetcenter.net/iau/lists/Centaurs.html}}.  Red diamonds in Figure \ref{ea} represent  active Centaurs studied by \citet{2009AJ....137.4296J, 2012AJ....144...97G}. Blue circles represent our HPO sample. Four black curves show the loci of orbits having perihelion distances equal to the aphelia of Jupiter, Saturn, Uranus and Neptune, marked $J$, $S$, $U$ and $N$, respectively. Figure \ref{qi} shows the perihelion versus  inclination plane, with the same color-coding as Figure \ref{ea}.  These two figures show that our  sample occupies a wide range of semimajor axes and inclinations, consistent with the Neptune-scattered objects, all with perihelion distances larger than 15 AU, but smaller than 30 AU.   \\

\subsection{Photometry}

\begin{figure}
\epsscale{1.0}
\plotone{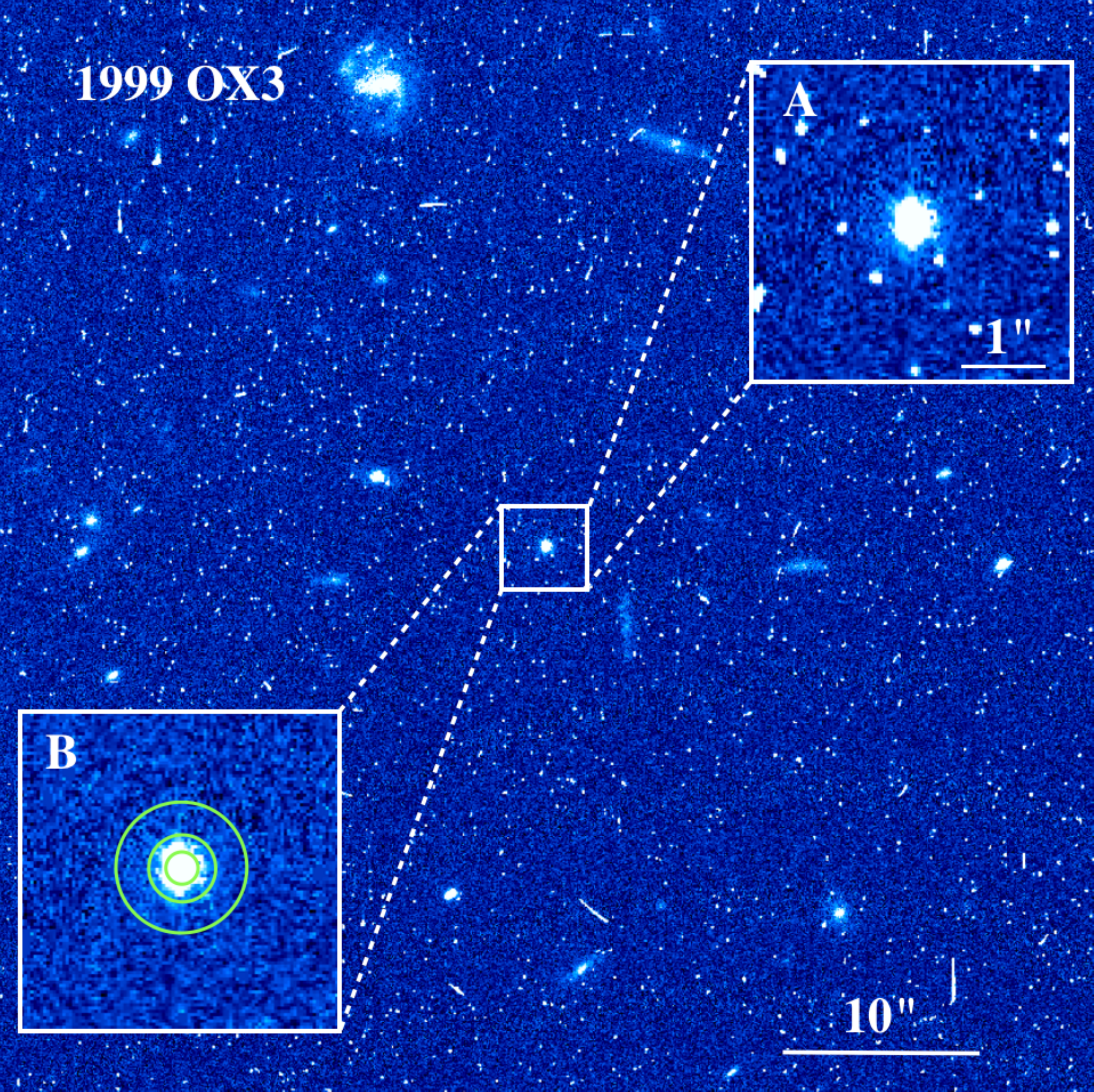}
\caption{Single 300 s integration HST image of 1999 OX3 at $r_H$ = 18.5 AU, shown with a logarithmic stretch between data numbers -0.08 and 0.15. Panel A is a zoom-box to show the immediate vicinity of the Centaur.  Panel B shows the same region as Panel A, cleaned of cosmic rays, and with green circles marking the 0.2\arcsec, 0.4\arcsec~and 0.8\arcsec~radius (corresponding to 5, 10 and 20 pixels) photometry apertures.
\label{images}}
\end{figure}

Cosmic rays are abundant in HST data, and so we developed a scheme for their removal. First, the two images of each object were registered and then subtracted from each other. In the resulting difference images, most pixels have values close to zero except where cosmic rays fall in either image. Next, we determined the distribution of pixel differences  and applied a sigma cutoff to identify cosmic ray contaminated pixels, which we replaced with the local average value. The entire image plane was treated in this way, except that  a circular region 10 pixels in radius and centered on the object was excluded  in order to prevent accidental replacement of real signal. Such special treatment can leave cosmic rays within the central area unaffected. In these rare cases, and where possible, we manually removed the cosmic rays from the area by interpolation. Figure \ref{images} shows a sample image, with inset boxes showing the region around the target both before and after cosmic ray removal. More than 70\%~of the targets have two useful observations and we present the average of the two measurements in  this paper.    In the remaining cases, cosmic rays unfortunately overlap the target  in one of the two images, but in no case were both images so affected. Where necessary, we report data using only a single image. As a test, we compared photometry from the automatically cosmic ray cleaned images with  photometry from images in which cosmic rays were removed manually, finding no significant differences. We used the photometry from the automated technique in our analysis.  \\

%\subsection{Sample Orbital Parameters}

We successfully observed 53 HPOs.  The brightness of the nucleus was measured using an extraction aperture 0.2\arcsec~(5 pixels, see the innermost green circle in Figure \ref{images}) in radius. The sky background was determined in a concentric ring with inner and outer radii 0.8\arcsec~and 4.0\arcsec~(20-100 pixels). The outmost green circle is the 20-pixel radius in Figure \ref{images}. We converted the apparent magnitudes from instrument magnitudes assuming HST calibration to a solar-type (G2V) spectrum. Table (\ref{tobs}) lists the observed objects with their observing date UT, heliocentric and geocentric distances, phase angle, and measured apparent magnitude ($V$). \\

%Figure (\ref{compare-appmag}) compares the apparent magnitudes, $V$, with predicted values listed on the JPL Horizons site, $V_{JPL}$. The solid straight line shows $V = V_{JPL}$.  A majority of the target objects are fainter than predicted by Horizons, in the sense $V_{JPL} = V - 0.2$ (dashed line in Figure \ref{compare-appmag}).   This systematic offset is likely a result of both an uncertain color term in transforming the broadband F350LP magnitudes to $V$.
%%, and the phase function correction employed by the Horizons software to predict the magnitude at the times of the HST observations.  For example, the Horizons site uses a phase function from Bowell et al.~(1989) with angle parameter $G$ = 0.15, which gives a  phase correction of 0.31 magnitudes at $\alpha$ = 3\degr.  
%Remaining scatter about the dashed line is  nearly symmetric and presumably caused by different objects having different phase functions (Bauer et al.~2003), and  by rotational variations in the presented cross-sections, caused by aspherical shape.  For a few objects  the displacement from the dashed line is too large to be plausibly explained in this way, and the cause remains unknown.  %The two labelled in the FIGURE are the ones we should mention here
 
\begin{figure}
\epsscale{1.0}
\plotone{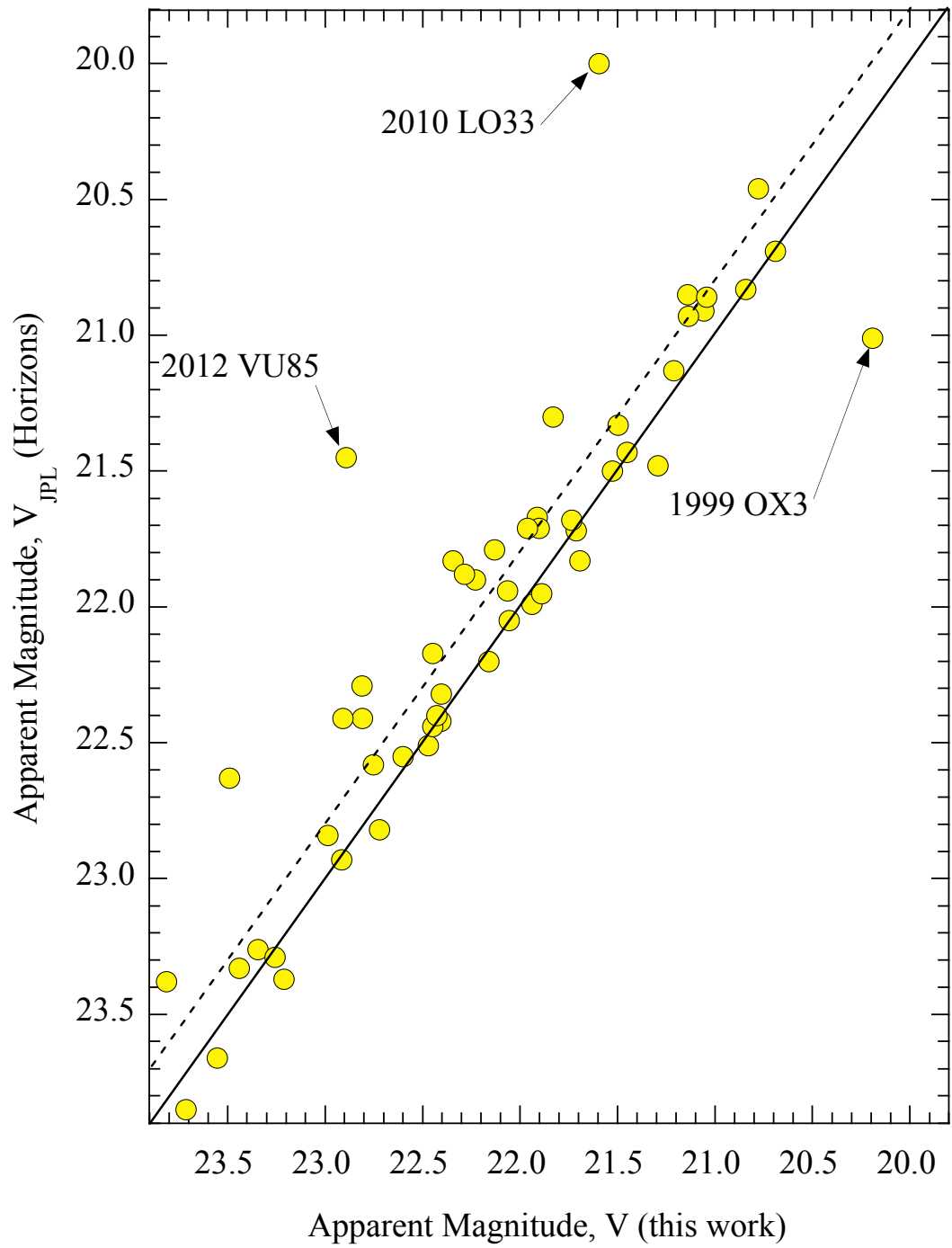}
\caption{Comparison of apparent magnitudes, $V$, from the current work with those from the JPL Horizons database, $V_{JPL}$.  The solid line shows equal magnitudes $V = V_{JPL}$.  The dashed line shows $V = V_{JPL} + 0.20$.
\label{compare-appmag}}
\end{figure}

Figure \ref{compare-appmag} compares the apparent magnitudes, $V$, with predicted values listed on the JPL Horizons site, $V_{JPL}$. The solid straight line shows $V = V_{JPL}$.  A majority of the target objects are fainter than predicted by Horizons, in the sense $V = V_{JPL} + 0.2$ (dashed line in Figure \ref{compare-appmag}).   This systematic offset is likely a result of  a  color term implicit  in transforming the very broadband F350LP magnitudes to $V$.
%, and the phase function correction employed by the Horizons software to predict the magnitude at the times of the HST observations.  For example, the Horizons site uses a phase function from Bowell et al.~(1989) with angle parameter $G$ = 0.15, which gives a  phase correction of 0.31 magnitudes at $\alpha$ = 3\degr.  
Remaining scatter about the dashed line is  nearly symmetric and presumably caused by different objects having different phase functions \citep{2013ApJ...773...22B}, and  by rotational variations in the presented cross-sections, caused by aspherical shape.  For a few objects (notably 1999 OX3, 2010 LO33 and 2012 VU85) the displacement from the dashed line is too large to be plausibly explained in this way, and the cause remains unknown. 
%The two labelled in the FIGURE are the ones we should mention here

%\section{Discussion}\label{sec:dis}

%\setcounter{table}{2}

\begin{figure}
\epsscale{1.0}
\plotone{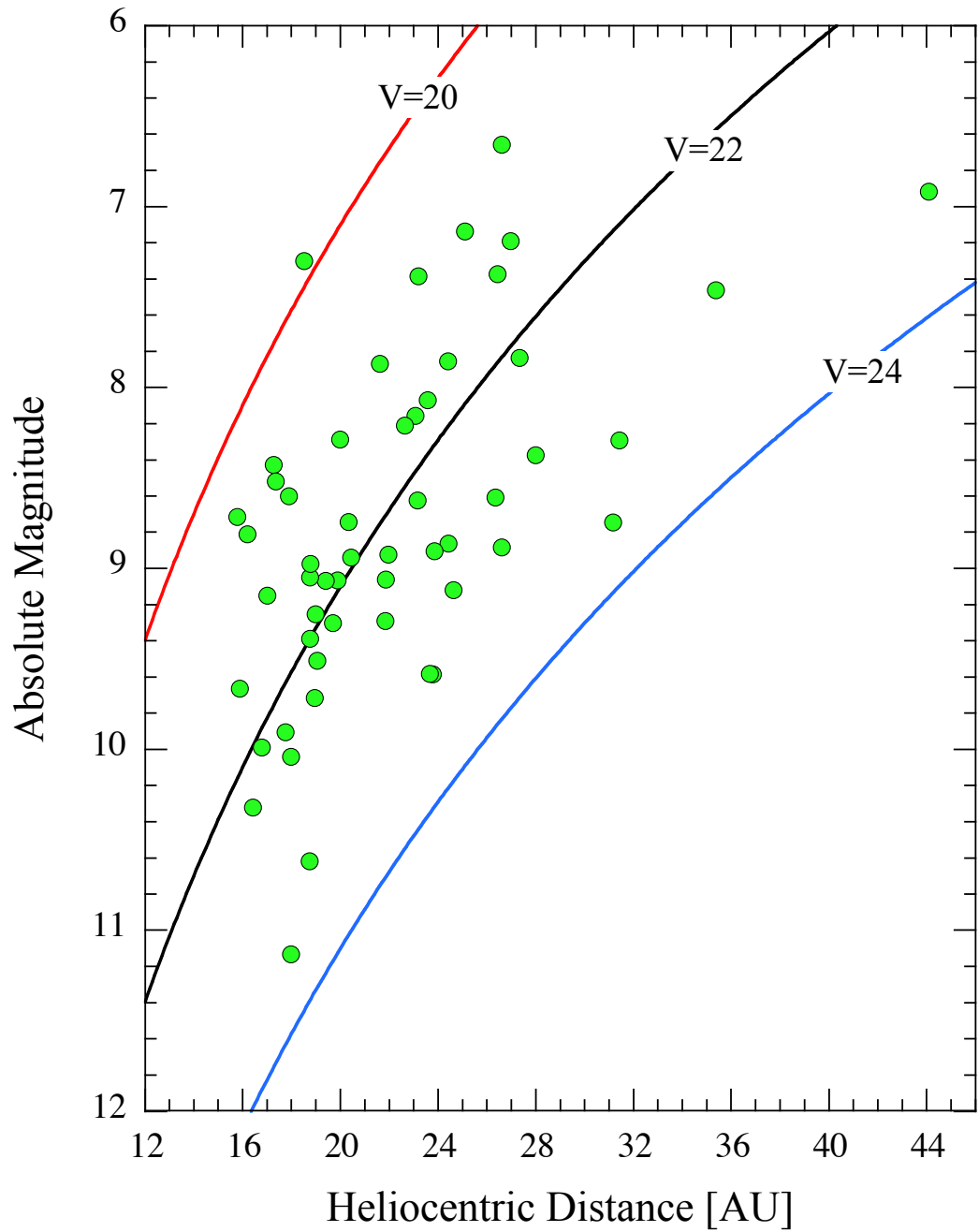}
\caption{Absolute magnitude of the High Perihelion Object (HPO) sample as a function of heliocentric distance.  Lines show the apparent magnitudes, as labeled, for opposition observations (i.e.~$\Delta = r_H - 1$, and $\alpha$ = 0\degr).  
\label{absmag-distances}}
\end{figure}

%\subsection{Absolute Magnitudes}
We converted the apparent magnitudes to absolute magnitudes using

\begin{equation}
H_V= V - 5\log_{10}(r_H \Delta) + 2.5\log_{10}(\Phi(\alpha))\label{H}
\label{H}
\end{equation}

\noindent where $\Phi(\alpha) \le 1$ is the phase function, equal to the ratio of the light scattered at phase angle $\alpha$ to that at $\alpha$ = 0\degr, and $r_H$ and $\Delta$ are the heliocentric and geocentric distances, respectively, both  in AU.  The phase functions of Centaurs reported by \citet{2003Icar..166..195B} show a wide range, with $G$ parameters from -0.13 to +0.18 in the ``HG'' system defined by \citet{1989aste.conf..524B}.  Similar diversity was reported by \citet{2007AJ....133...26R}, who also noted that the phase functions of Centaurs tend to be linear with $\alpha$.  Additional phase function measurements, also showing a wide scatter, are given in Alvarez-Candal et al.~(2016) and Ayala-Loera et al.~(2018).  Motivated by these results, and in the interests of simplicity, we write $2.5\log_{10}(\Phi(\alpha)) = -\beta \alpha$, where $\beta$ = 0.061$\pm$0.002 magnitudes per degree is  the mean value of the V-filter phase coefficients reported for seven Centaurs in Table 4 of Rabinowitz et al.~(2007).   Given that the maximum phase angles attained by the Centaurs in our sample are $\alpha_{max} \sim 3\degr$ (Table \ref{tobj}), the necessary phase corrections are modest ($\beta \alpha_{max}$ = 0.12 magnitudes). Uncertainties in the value of $\beta$ render the derived absolute magnitudes uncertain by a similar amount.   This is large compared to the photometric errors of the HST data, but comparable to or smaller than the likely variations in absolute magnitude caused by asphericity and rotation of the Centaurs. The apparent ($V$) and  absolute magnitudes ($H$)  are found in Tables (\ref{tobs}) and  (\ref{tquantities}), respectively. The absolute magnitudes  as functions of the heliocentric distances are shown in Figure \ref{absmag-distances}. As an example, 1999 OX3 is studied by others researchers. Our measurement, $H_V=7.3$, is comparable to $H_R=7.1$ reported by \citet{2003Icar..166..195B}, but brighter than $H_V=7.980\pm0.092$, and $7.60\pm0.06$  measured by \citet{2016A&A...586A.155A,2018MNRAS.481.1848A}.\\

\subsection{Effective Radii}
%For each Centaur, we also calculate the nucleus cross-section, $C_n$, from Equation (\ref{invsq}) by substituting $V_1$ for $V_c$, and the radius of an equal-area circle from $r_n = (C_n/\pi)^{1/2}$.  These are listed in the last column in Table (\ref{tobj}). The diameters of Centaurs are compared with their perihelion distance in Figure (\ref{radii-q}). 

To convert the magnitudes to effective radii, $r_e$, we use the relation

\begin{equation}
r_e = \left(\frac{1.5\times10^{8}}{p_V^{1/2}}\right) 10^{0.2(V_{\odot} - H)}
\label{r_e}
\end{equation}

\noindent in which $1.5\times10^{8}$ is the number of kilometers in 1 AU, $V_{\odot}$ = -26.74 is the V-band magnitude of the Sun and $p_V$ is the geometric albedo. \\

\begin{figure}[h!]
\epsscale{1.0}
\plotone{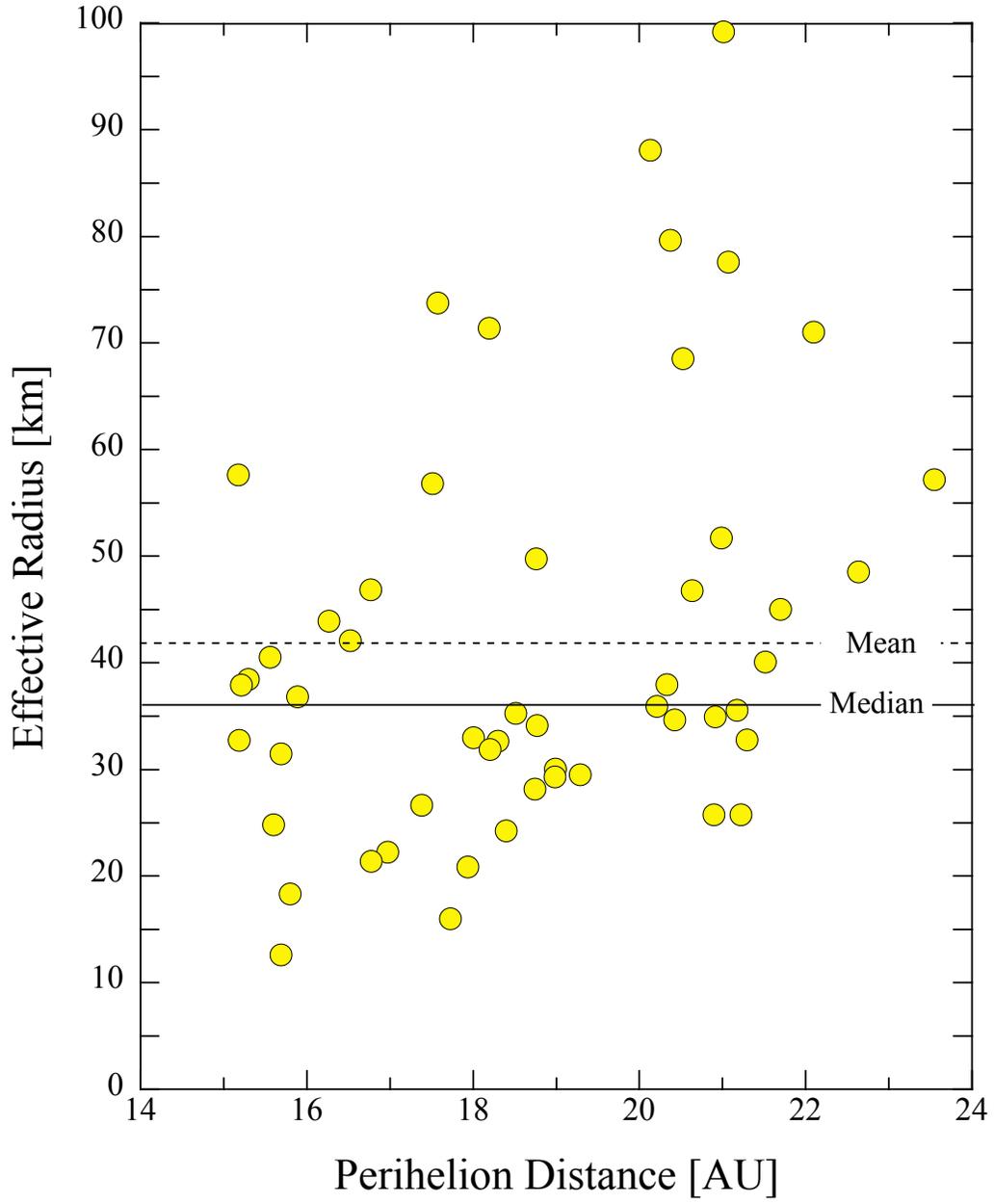}
\caption{Effective radii, $r_n$, as a function of perihelion distance, $q$. Assuming $p_V=0.1$, the median radius is $r_n$ = 36 km (solid line) while the average radius is 42 km (dashed line).\label{radii-q}}
\end{figure}

As is true of the phase functions, the geometric albedos of most HPOs are unmeasured.   The mean albedo of 52 Centaurs and Scattered KBOs is $p_V = 0.08\pm0.04$ studied by \citet{2013ApJ...773...22B}, while the mean albedo of 16 Centaurs (in a largely over-lapping object sample) reported  by \citet{2014A&A...564A..92D}  is $p_V = 0.07\pm0.05$.  The former authors also reported a color-albedo dependence, with ``blue'' objects having $p_V = 0.06\pm0.02$ and ``red'' objects $p_V = 0.12\pm0.05$ (c.f.~Figure 3 of \citet{2014ApJ...793L...2L}). However, it is worth noting that albedo determinations are difficult and the quoted uncertainties in some cases appear to underestimate the true uncertainties in the reported albedo.  For example, 250112 (2002 KY14) was reported by \citep{2013ApJ...773...22B}, to have $p_V = 0.185\pm0.046$, a value that is three times larger than $p_V = 0.057_{-0.007}^{+0.011}$ as reported by \citet{2014A&A...564A..92D}.   For simplicity, in this work we assume a nominal  $p_V$ = 0.1 for all objects in order to evaluate object dimensions on a uniform basis.  Effective radii computed in this way can be easily scaled to other values of the albedo as they become available, in proportion to $(0.1/p_V)^{1/2}$ (Equation \ref{r_e}). The effective radii of the Centaurs, computed from their apparent magnitudes using Equations (\ref{H}) and (\ref{r_e}), are listed in Table (\ref{tquantities}) and plotted against the perihelion distance  in Figure \ref{radii-q}. Evidently, the radii of most objects in our sample fall in the  range 10 $\lesssim r_e \lesssim$ 100 km, with a sample median  $r_e$ = 36 km. We note that \citet{2003Icar..166..195B,2013ApJ...773...22B} gave the radii of Centaurs 1999 OX3 ($\sim$89 km), 2002 CR46 ($85\pm35$ km) and 2002 XU93 ($96\pm 35$ km), all of which are comparable to our results  (74, 57, and 52 km, respectively) within the uncertainties. \\

\subsection{Coma Detections}
\begin{figure}
\epsscale{1.0}
\plotone{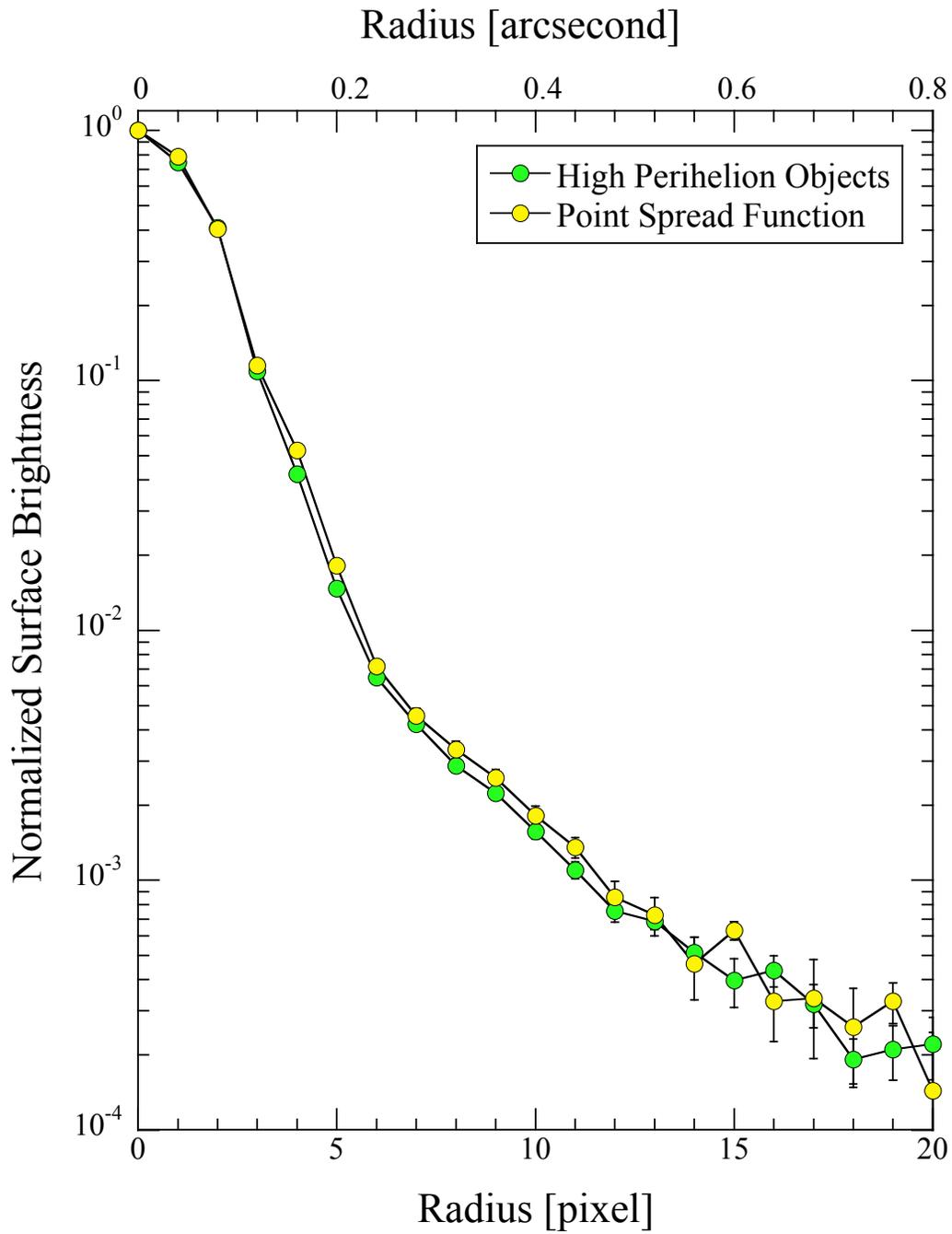}
\caption{Normalized mean radial surface brightness profiles of  High Perihelion Objects (HPOs, green filled circles) and stars (yellow filled circles) with error bars computed from the standard deviations of 53 object and 27 stellar profiles.
\label{radprof}}
\end{figure}

We search for coma by comparing annular photometry of the HPOs with reference stars measured identically. Stars in our HPO dataset are all trailed by the non-sidereal motion of HST.  Instead, to obtain an empirical measure of the point-spread function (PSF) we used the profiles of  reference stars in untrailed (i.e.~sidereal target) archival WFC3 images taken through the F350LP filter with comparable integration times. We used the HST archive to identify 27 stars free from confusion with nearby stars and galaxies, and which have  apparent magnitudes similar to those of the HPO sample. The average surface brightness  profiles derived from the stars and the HPOs are compared in Figure \ref{radprof}, showing good agreement within the uncertainties of measurement.  The HST observations provide no evidence that the HPOs, as a group, exhibit coma.\\

Next, we use aperture photometry as the primary means to quantify the presence of coma in individual objects. This technique is robust, simple to apply and to interpret and, given the narrow and  stable point spread function of the HST, is also very sensitive.  A significant advantage is that the technique  is independent of the morphology assumed by the ejected dust, whether it be spherically symmetric or highly collimated into a tail or trail by the action of solar radiation pressure.   We base our analysis on \citet{2009AJ....137.4296J}, with small modifications described below.  \\

All targets and point sources (stars) were measured using three concentric apertures of radii (see green circles in Figure \ref{images}) $\theta_0=0.2$\arcsec~ (5 pixels), $\theta_1=0.4$\arcsec~ (10 pixels) and $\theta_2=0.8$\arcsec~ (20 pixels), yielding the three apparent magnitudes, $V_0$, $V_{1}$ and $V_{2}$.  Note that $V_0$ is the nucleus magnitude. \\

We define the annular magnitude excess  as

\begin{equation}
\Delta  V_{i,j}= (V_i-V_j)_{HPO} - \overline{(V_i-V_j)_{\star}}
\label{vc}
\end{equation}

\noindent where   the aperture pairs are $i,j$ = (0, 1), (0, 2), and (1, 2). Subscripts ``HPO'' and ``${\star}$'' refer to the high perihelion object and the star profiles.   The average, median and standard deviations of quantities $(V_i-V_j)_{HPO}$ and $(V_i-V_j)_\star$ are listed in Table (\ref{tmagdiff}). The mean and median quantities are equal within the uncertainties. Note that all entries are consistent with $\Delta V_{i,j}$ = 0 at the  3$\sigma$ level.   For reference, we also measured in the same way a synthetic image created using the TinyTim simulation software (Krist et al.~2011).  This image gave results slightly different from the empirical determinations listed in Table (\ref{tmagdiff}) (e.g.~($V_1 - V_2)_{\star}$ = 0.034 vs.~0.048 from the Table) and we elected to use the empirical determinations over those from TinyTim.\\

%JING - TABLE 4 LOOKS WRONG.  Angles wrong, delta V wrong.  Please check.

\begin{figure}[h!]
\epsscale{1.0}
\plotone{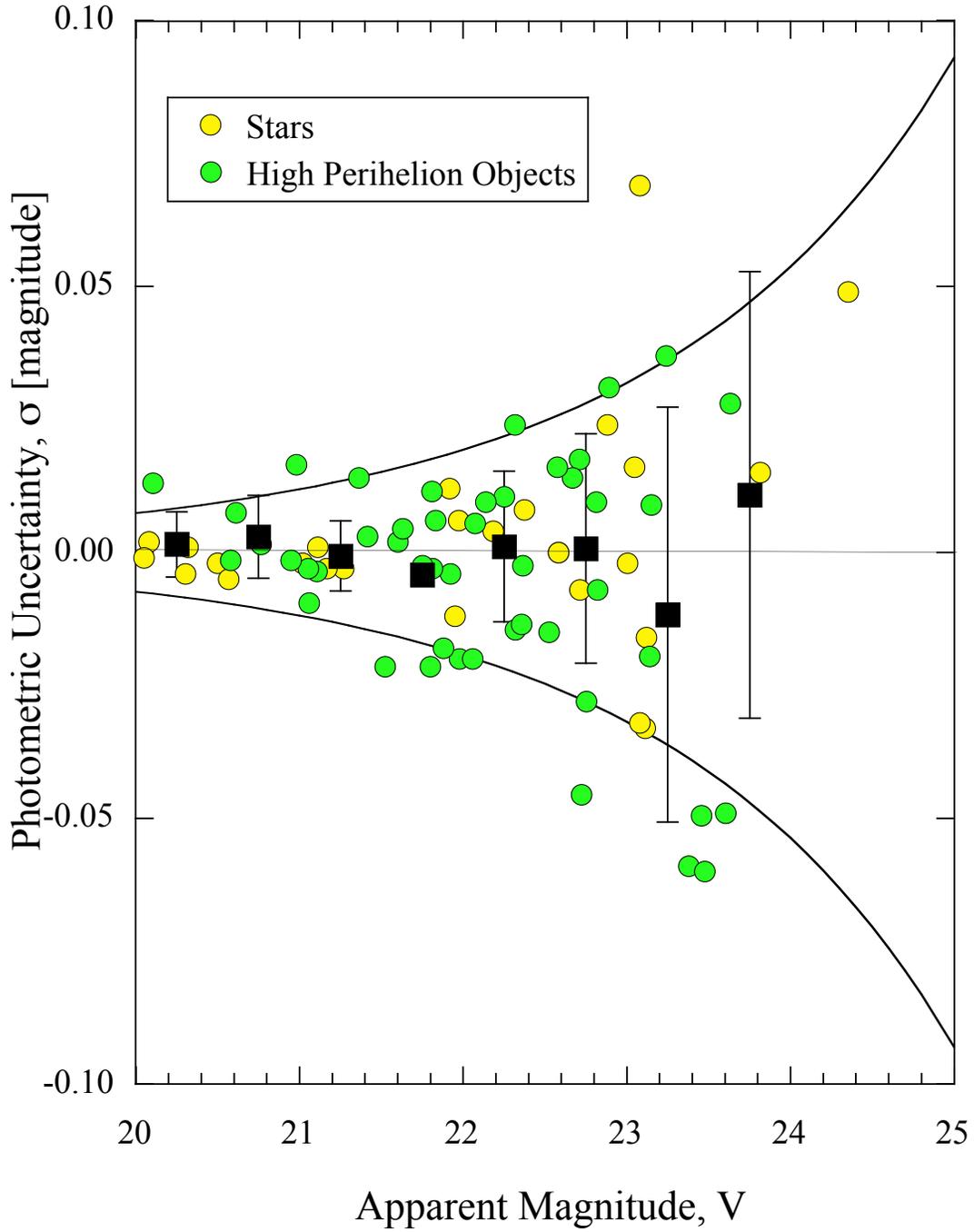}
\caption{Uncertainties in the measured annular magnitudes as a function of the source brightness.  Yellow and green symbols distinguish HPOs from field stars.  The curved lines show the expected noise based on a model.  Grey squares show the mean and $\pm$1 standard deviation within a series of apparent magnitude bins each 0.5 magnitude wide.  The concordance between the actual scatter and the expected uncertainties shows that the WFC3 data are photometrically well-behaved.  \label{sigmaplot}}
\end{figure}

Figure \ref{sigmaplot} shows $\Delta V_{1,2}$ as a function of the $V$ magnitude for stars (yellow circles) and HPOs (green circles).  For the stars, we replaced $(V_i-V_j)_{HPO}$ in Equation (\ref{vc}) with the individual aperture difference for each star.  Grey squares with error bars show the mean and standard deviation of $\Delta V_{1,2}$ within a series of magnitude bins each 0.5 magnitude wide.  The scatter of the measurements clearly grows as the objects become fainter, from $\pm$0.006 magnitudes at $V$ = 20.25 to $\pm$0.040 at $V$ = 23.75.  Curved lines in the figure show a model of the uncertainty incorporating both photon (Poisson) noise and CCD readout noise.  For the latter, we assumed the canonical WFC3 values for read noise = 3 electrons, digitization at 1.5 electrons per ADU, and that a V = 0 source gives 4.72$\times10^{10}$ ADU s$^{-1}$.  We also used the measurement, from stars, that  4.8\% of the light from a point source falls in the annulus between 10 and 20 pixels radius.  The agreement between the model and actual uncertainties gives us confidence that the data are essentially photon limited, and therefore of very high quality.  \\

\begin{figure}
\epsscale{1.0}
\plotone{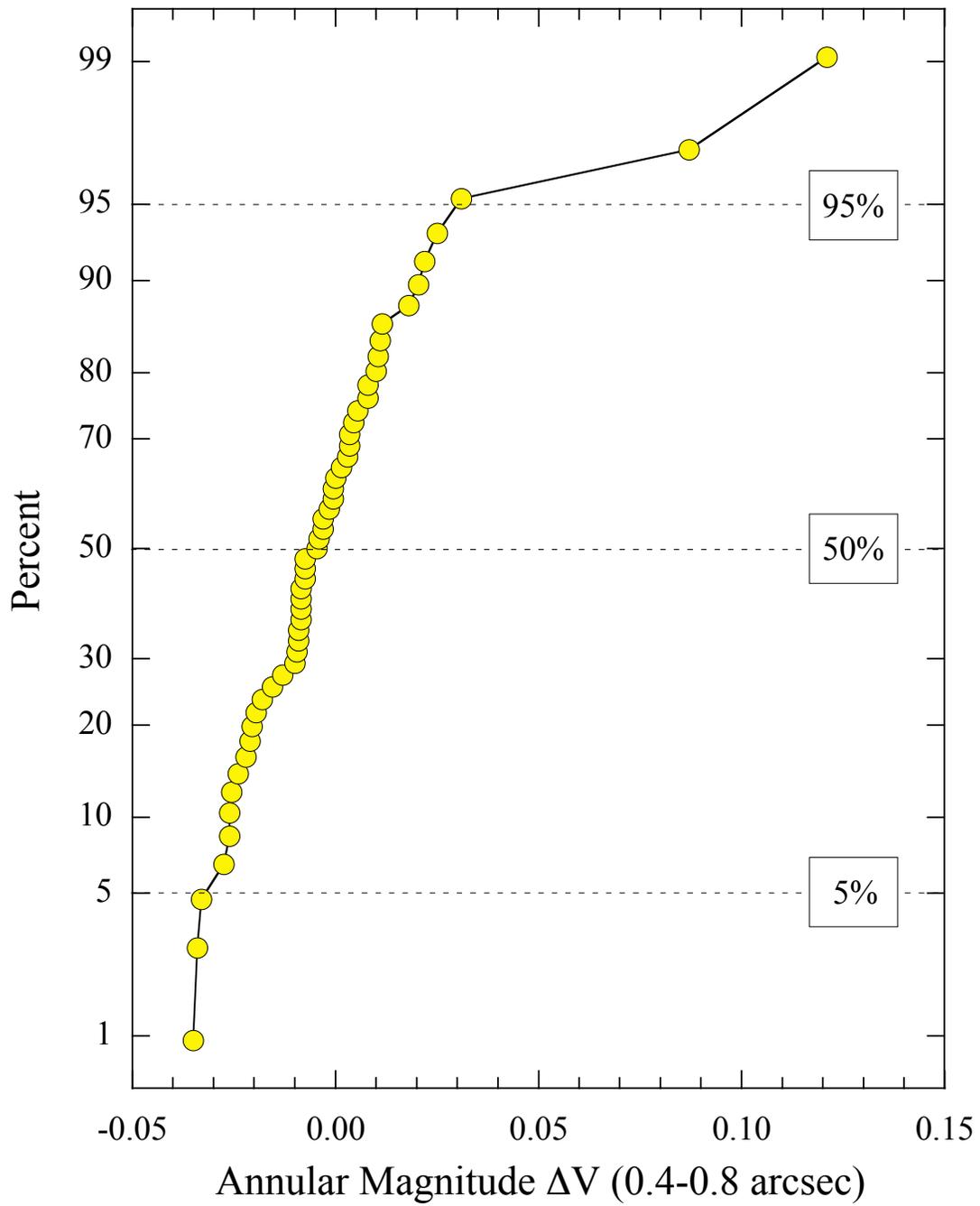}
\caption{Cumulative distribution of annular magnitudes $\Delta V_{1,2}$ ($\theta =0.4$\arcsec~to 0.8\arcsec).  Dashed horizontal lines mark the 5\%, 50\% and 95\% values, as labelled.  \label{vc-histogram}}
\end{figure}

The cumulative distribution of $\Delta V_{1,2}$ is displayed in Figure \ref{vc-histogram}, where $ \overline{(V_i-V_j)_{\star}}=0.051$ (see Table \ref{tmagdiff}). All but two HPOs have  $\Delta V_{1,2}<0.03$. The $\Delta V_{1,2}$ Gaussian best-fit distribution has a $\sigma=0.042$  centered at -0.007. Within the uncertainties, the brightness enhancement around the HPOs  is zero, indicating that comae of HPOs are not detected.  \\

\subsection{Mass Loss Rates}
We derive upper limits to the mass loss   from measured upper limits to the coma. The mass of a collection of spherical  particles can be  written as

\begin{equation}
M= \int_{a_0}^{a_1} \frac{4}{3}\pi \rho a^3 n(a) da.
\label{m}
\end{equation}
\noindent Here, $\rho$ is the particle density, and $n(a)da$ is the differential size distribution, equal to the number of particles with radii between $a$ and $a + da$, and distributed in the range  $a_0 \le a \le a_1$. It is convenient to assume a power-law size distribution, $n(a)da = \Gamma a^{-\gamma} da$, where $\Gamma$ and $\gamma$ are constants.  The total cross-section of the particles is

\begin{equation}
C = \int_{a_0}^{a_1} \pi a^2 n(a) da.
\label{c}
\end{equation}

\noindent Equations (\ref{m}) and (\ref{c}) can be combined to give

\begin{equation}
M = \frac{4}{3} \rho \overline{a} C
\label{mass}
\end{equation}

\noindent for all $\gamma \neq$ 3 or 4, where  $\overline{a}$ is the average particle radius.   We adopt $\gamma$ = 3.5, based on measurements of comets \citep{2001A&A...377.1098G}, for which   $\overline{a} = \sqrt{a_0 a_1}$.  We adopt $a_0$ = 0.1 $\mu$m, since smaller particles are inefficient optical scatterers and contribute little to the scattered light intensity.  The appropriate value of $a_1$ is less certain.  Since we are mainly interested in scaling from one object to the next, we adopt  $a_1$ = 1 mm, giving $\overline{a}$ = 10 $\mu$m.  Then, Equation (\ref{mass}) allows us to calculate the particle mass corresponding to the cross-section  inferred from the photometry.  \\

Equation (\ref{mass}) gives the particle mass projected within an annulus around the nucleus. The approximate residence time of particles in the annulus is

\begin{equation}
\tau= \frac{\Delta r}{U}
\label{dt}
\end{equation}

\noindent where $U$ [m s$^{-1}$] is the average speed of the particles leaving the nucleus.  The width, $\Delta r$, is  related to the  angular dimensions of the annulus by

\begin{equation}
\Delta r = s \delta \theta  \Delta 
\label{dr}
\end{equation} 

\noindent where $\delta \theta = \theta_j-\theta_i$ is in arcseconds, $\Delta$ is in AU, and scale factor $s = 7.25\times 10^5$ m arcsec$^{-1}$ AU$^{-1}$  is the number of meters in one arcsecond at one AU.  \\
%%%%

The particle speed, $U$, in Equation (\ref{dt}) is unmeasured.  Practical lower and upper limits to $U$ are set by the gravitational escape speed and the gas sound speed, respectively.   The escape speed, for a sphere of radius $r_e$ and density $\rho_n$ is given by

\begin{equation}
V_e = \left(\frac{8 \pi G \rho_n}{3}\right)^{1/2} r_e
\label{v_e}
\end{equation}

\noindent where $G = 6.67\times10^{-11}$ N kg$^{-2}$ m$^2$ is the gravitational constant.  With $\rho_n$ = 1000 kg m$^{-3}$ and $r_e$ in [km], Equation (\ref{v_e}) gives $V_e = 0.75 r_e$ [m s$^{-1}$].   At the median radius $r_e =$ 36 km, the typical escape speed is $V_e$ = 27 [m s$^{-1}$].\\

The gas sound speed is

\begin{equation}
V_{th} = \left(\frac{8 k T}{\pi \mu m_h} \right)^{1/2}
\label{v_th}
\end{equation}

\noindent where $k = 1.38\times10^{-23}$ J K$^{-1}$ is Boltzmann's constant, $T$ is the gas temperature, $\mu$ is the molecular weight of the gas and $m_H = 1.67\times 10^{-27}$ kg is the mass of the hydrogen atom.  As previously noted, water ice does not sublimate appreciably at the large heliocentric distances considered here and only supervolatiles constitute plausible sources of gas drag.  We solved the energy balance equation for sublimating CO ($\mu$ = 28) to find that the equilibrium sublimation temperature is negligibly dependent on heliocentric distance  (we find $T$ = 26.6 K at $r_H$ = 5 AU falling to $T$ = 24.4 K at 30 AU) as a result of the dominance of the latent heat term in the energy balance.  We take $T \sim$ 25 K, giving $V_{th}$ = 120 m s$^{-1}$.  This is consistent with thermal broadening of CO rotational lines in distant bodies (e.g. \citet{1994Natur.371..229S} measured a  CO line width of $\sim$200 m s$^{-1}$ in 29P when at $r_H$ = 6 AU, while \citet{1995Icar..115..213C}  independently measured 140 m s$^{-1}$ in the same object).  \\

Even for the largest objects in our sample (see Table \ref{tquantities}), we find $V_{th} > V_e$, consistent with gas drag expulsion of particles\footnote{As an aside, we note from Equations (\ref{v_e}) and (\ref{v_th}) that $V_{th} = V_{e}$ when $r_e =$  160 km.  In all larger objects, gas drag will be unable to eject even the smallest particles against self-gravity and only sub-orbital dust trajectories will result.  The scattering cross-section and time-dependent brightness of such a body could still be affected by dust but no resolved coma would be present. None of the objects in our sample are this large.}.    Finally, the mass loss rate, $\dot{M} \sim M/\tau$, is obtained by combining Equations (\ref{mass}), (\ref{dt}) and (\ref{dr}) to find

\begin{equation}
\frac{dM}{dt}=\frac{4}{3}\left(\frac{\rho\overline aC_{}}{s\delta \theta\Delta}\right) U 
\label{dmbdt}
\end{equation}

\noindent~
Since $\dot{M} \propto U$, and we are interested in setting upper limits to the mass loss, we set $U = V_{th}$ = 120 m s$^{-1}$ in Equation (\ref{dmbdt}). The resulting mass loss rates are listed in the last column in Table (\ref{tquantities}). In the table, centaurs are listed in order of their designated names. \\

\section{DISCUSSION}

\begin{figure}
\epsscale{1.0}
\plotone{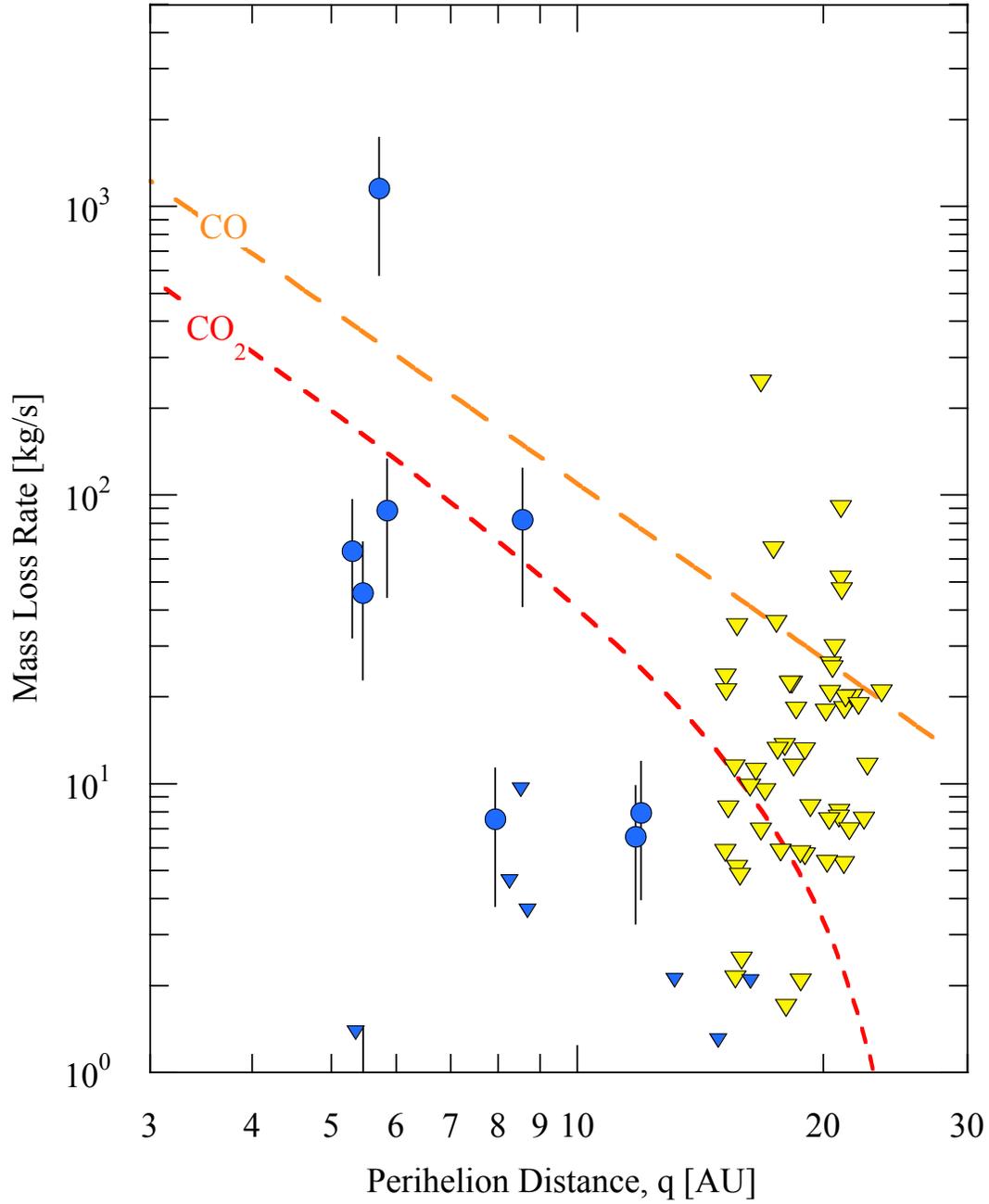}
\caption{Mass loss rate vs.~perihelion distance.  Yellow filled triangles indicate upper limits from this work.  Blue symbols denote data from Jewitt (2009), with circles indicating detections and triangles indicating upper limits.  Error bars on the blue circle detections are added to represent factor of two uncertainties.  The orange and red dashed lines show solutions to Equation (\ref{energy}) for  CO and CO$_2$ ices, respectively, with exposed surface areas of 5 km$^2$. \label{dmbdt_vs_q2}}
\end{figure}

Figure \ref{dmbdt_vs_q2} shows the derived mass loss rates as a function of the perihelion distance, with downward-pointed, yellow-filled triangle symbols used to indicate upper limits.  The figure shows that no activity was detected in any of the HPOs observed in the present study.  The (model-dependent) mass loss limits range from $\sim2$ kg s$^{-1}$ to $\sim$10$^2$ kg s$^{-1}$, depending on the size and distance to each object.   The median limiting mass loss rate in the HPO sample is $\dot{M} <$  11 kg s$^{-1}$ and the median perihelion distance $q$ = 19 AU.   \\

For comparison, we include in Figure \ref{dmbdt_vs_q2} estimates of $\dot{M}$ from \citet{2009AJ....137.4296J}  obtained using ground-based telescopes on a selection of Centaurs having (mostly) smaller perihelion distances than in our current sample.  In order to permit direct comparison with the HST measurements we have scaled the $\dot{M}$ values from column 9 of their Table 4 to the same particle size $\overline{a}$ = 10 $\mu$m as used here.  Non-detections are again indicated by (blue) downward pointing triangles while coma detections are marked with blue-filled circle symbols and error bars equal to a factor of two in the production rates.    The median  mass loss rate  and perihelion distance of the active Centaurs are  46 kg s$^{-1}$ and 5.8 AU, respectively. Other measurements of Centaurs reported by \citet{2019A&A...621A.102C} refer to objects close to the limiting magnitude of their survey.  They have negligible sensitivity to extended emission from outgassing and are not considered here.  \\

Five of the eight  active Centaurs in the Figure exhibit mass loss rates large enough to have been detected in the HPO sample, had activity been present.  In this sense, the non-detection of activity is consistent with the crystallization hypothesis, which predicts that no examples of activity should be found.  Of course, while consistent, we cannot argue that the new data prove the crystallization hypothesis.  This is because the sensitivity to coma achieved by HST is insufficient to rule out the presence of low-level supervolatile sublimation at 15 AU and beyond.  \\

To examine the nature of low-level activity at large $r_H$, we consider the  energy balance  for a sublimating surface, neglecting conduction, expressed as

\begin{equation}
\frac{L_{\odot}(1-A)}{4\pi r_H^2} = \chi\left[\varepsilon \sigma T^4  + f_s(r_H) L(T)\right].
\label{energy}
\end{equation}

\noindent Here, $A$ and $\varepsilon$ are the Bond albedo and emissivity of the sublimating surface, $L_{\odot}$ is the solar luminosity, $r_H$ is heliocentric distance expressed in meters, $\sigma$ is the Stefan-Boltzmann constant and $L(T)$ is the temperature-dependent latent heat of sublimation.  Quantity $f_s$ [kg m$^{-2}$ s$^{-1}$] is the sublimation mass flux.  We assume $A$ = 0.04, $\varepsilon$ = 1 while noting that solutions to Equation (\ref{energy}) are insensitive to both quantities provided $A \ll 1$ and $\varepsilon \gg$ 0.  Parameter $\chi$ is a dimensionless number that expresses the distribution of absorbed energy over the nucleus, varying between $\chi$ = 1 for a flat surface oriented perpendicular to the Sun-comet line and $\chi$ = 4 for an isothermal sphere.  We adopt $\chi$ = 2 as the intermediate case, corresponding to hemispheric warming of a spherical nucleus.   \\

We solved Equation (\ref{energy}) using thermodynamic parameters for CO and CO$_2$  ices tabulated by \citet{brown-ziegler1980}.  The results are plotted in Figure \ref{dmbdt_vs_q2} where, for reference, we show the mass sublimated per second from an exposed 5 km$^2$ area of each ice (CO in orange, CO$_2$ in red).  As noted above, the sublimation of CO declines closely as $r_H^{-2}$ across the planetary region of the solar system while CO$_2$ exhibits a downturn beginning at about 15 AU.  \\

%The correspond to sublimation from exposed areas of CO, $C_{CO} = \dot{M}/f_s$, in the range  0.2 $\le C_{CO} \le$ 20 km$^2$ while, for CO$_2$, areas larger by about an order of magnitude are implied.  Compared to the surface area of a sphere having the median HPO radius, $r_e$ = 35 km, the active fraction is limited to $10^{-5} \le C_{CO}/(4\pi r_e^2) \le 10^{-3}$.  \\
%%%%

The observations set strong limits on the possible area of exposed supervolatiles on each object.   For example, at the median perihelion distance $r_H$ = 19 AU, the equilibrium sublimation flux of CO from a flat surface oriented perpendicular to the Sun-object line is $f_s = 6\times10^{-6}$ kg m$^{-2}$ s$^{-1}$ (Equation \ref{energy}).   Rate $\dot{M}$ then  corresponds to sublimating area $A = f_s^{-1} \dot{M}$, assuming a gas to dust mass production rate of unity.  The empirical median rate $\dot{M} <$  11 kg s$^{-1}$ would reflect sublimation from an area $A < 2$ km$^2$.  The implied active fraction is $f_A = A/(4\pi r_e^2)$, where $r_e$ = 36 km is the median radius from Table (\ref{tquantities}). We obtain $f_A < 10^{-4}$, which is two to four orders of magnitude  smaller than $f_A$ measured on the nuclei of active Jupiter family comets \citep{1995Icar..118..223A}.  Small values of $f_A$ are to be expected since exposed CO sublimates rapidly even at 19 AU ($f_s/\rho \sim$ 20 cm year$^{-1}$, or several meters per orbit). \\

We briefly discuss the distance-dependence of bias effects in imaging faint comae.  First, the rate of supervolatile sublimation varies as $r_H^{-2}$ (Figure \ref{dmbdt_vs_q2}).  Scaling from median distance $q$ = 5.8 AU to 19 AU, for example, corresponds to a sublimation rate smaller by a factor of $\sim$11.  Second, the dust comae are observed in reflected sunlight, the intensity of which also falls as $r_H^{-2}$, together giving an $r_H^{-4}$ variation in the surface brightness of a coma at fixed linear distance from an object sublimating in energy equilibrium with sunlight.  Scaling from 5.8 AU to 19 AU then corresponds to a fading of coma surface brightness by a factor $\sim$115.  This steep variation is partially offset by  the improved angular resolution of HST, which allows measurement of any coma closer to the nucleus source (where the surface brightness is intrinsically higher, for a given mass loss rate) than would be possible in ground-based, lower resolution observations.  Still, the  bias against coma detection at large $r_H$ remains, and  low-level outgassing in the HPOs due to steady sublimation of exposed supervolatiles might be present but go unobserved. \\

\subsection{Binary  Fraction:} 
Binaries are abundant in the Kuiper belt; $\sim$20\% of the cold classical KBOs are binary, as are $\sim$5\% of the dynamically hot populations \citep{2008ssbn.book..345N,2019TNObook..9.1}. Since the HPOs are products of the dynamically excited portion of the Kuiper belt, it is reasonable to expect to find a binary fraction similar to the $\sim$5\% measured in the hot population.  Currently,  two binary Centaurs are known ((42355) Typhon I Echidna \citep{2006Icar..184..611N}, and (65489) Ceto/Phorcys \citep{2007Icar..191..286G}).   
 In our sample of 53 objects, for example, the expected number of binaries would be $\sim$2.5. However, \citet{2014MNRAS.437.2297B} examined the disruptive effects of close planetary encounters on objects dipping into the giant planet region, finding that only about 10\% of binaries should resist disruption over the long term.   If applied to our sample of 53 objects, Brunini's fraction suggests that we should find an average of only 0.25 binaries.
 
The maximum separation for a binary to be stable against solar gravitational perturbations is given by the Hill  radius, $r_{Hill}$,

\begin{equation}
r_{Hill} = q \left(\frac{\rho_n}{3 \rho_{\odot}}\right)^{1/3} \left(\frac{r_n}{r_{\odot}}\right)
\end{equation}

\noindent where $q$ is the perihelion distance, $\rho_n = 1000$ kg m$^{-3}$ and $\rho_{\odot} = 1400$ kg m$^{-3}$ are the densities of the object and the Sun, respectively, and $r_n$ and $r_{\odot}$ are the radius of the Centaur and the Sun. 
With $q \gg 1$ AU,  the angle subtended by the Hill radius is just $\theta_{Hill} \sim r_{Hill}/q$, which is expressed in arcseconds as

\begin{equation}
\theta_{Hill} \sim 10\arcsec \left(\frac{r_n}{50\textnormal{~km}}\right),
\label{Hill}
\end{equation}

\noindent where  $r_n$ is radius in km, independent of distance.   In Equation (\ref{Hill}) we have  normalized to  $r_n$ = 50 km for convenience.   
The Nyquist-sampled (two pixel) WFC3 resolution of 0.08\arcsec~corresponds to $\sim 0.008~\theta_{Hill}$ for a 50 km radius object.  Most resolved Kuiper belt binaries occupy the central $\sim$10\% of the Hill sphere \citep{2019TNObook..9.1}, corresponding to separations $<$ 1\arcsec, and still about 10 times the HST resolution.    \\

We visually searched the Hill spheres of the target objects by comparing consecutive images to locate co-moving objects.  Owing to asymmetries in the faint wings of the PSF, the sensitivity to binary companions is a complicated function of the component brightness ratio, the angular separation and the position angle.  It is therefore impossible to set a uniform limit to the presence of binary companions.  Binaries with a large ratio of component brightnesses can easily escape detection and our search should therefore be understood to refer to binaries with equal-sized components, of which we detected none.  The non-detection of binaries in the HPO sample sets a 3$\sigma$ limit to the average binary fraction $<$8\%, consistent with expectations above.     %Assuming a Poisson counting distribution, the likelihood of detecting one binary object in a sample of 53 objects with this average value is $\sim$20\%. 

%This constitutes a lower limit, however, since Centaurs are on average closer than Kuiper belt objects, allowing small separation binaries that would go unobserved in the Kuiper belt to be detected.  These small separation binaries are also the ones most likely to resist being tidally split.

%A simple hypothesis is that amorphous ice drives the Centaur activity starting at $\sim$10 AU, which leads to burial or destruction of ultrared material that is thermodynamically stable at larger distances.  Crystallization is exothermic, and the rearrangement of bonds between water molecules leads to the expulsion of trapped molecules, potentially including supervolatile species.  The detections of CO in 29P and other near-Sun Centaurs, then, could be the product of crystallization of amorphous water ice.

%A related observation is that the optical colors of the short-period comets are different from those of the Kuiper belt objects from which they are thought to be derived.   Ultra-red material is present in the Kuiper belt but not in the comet population.  The transition occurs near $\sim$10 AU, corresponding to the orbit of Saturn, with Centaurs having smaller perihelia being depleted in ultra-red matter compared to more distant Centaurs and to objects in the Kuiper belt \citep{2002AJ....123.1039J,2009AJ....137.4296J,2015AJ....150..201J}.  This color change has been attributed to surface blanketing as cometary activity turns on at $\sim$10 AU \citep{2002AJ....123.1039J}.

\clearpage

\section{SUMMARY}
We used the Hubble Space Telescope to image 53 high perihelion objects at $\sim$0.08\arcsec~angular resolution.  Objects in our sample were selected to have perihelion distances $q >$ 15 AU, where sub-solar radiation equilibrium temperatures are $T \lesssim$ 100 K and amorphous water ice is too cold to crystallize on the orbital timescale.

\begin{enumerate}

\item No evidence for activity was found. We set upper limits to the mass loss, $\dot{M}$, from each object based on near-nucleus photometry and a simple model. At the median distance of our sample objects (19 AU), the median value is $\dot{M} <$ 11 kg s$^{-1}$, with an object to object  range 1 $\lesssim \dot{M} \lesssim$ 231 kg s$^{-1}$.   

\item The non-detection of activity in our sample is consistent with the hypothesis that activity observed in dynamically similar objects having smaller perihelion distances is caused by the crystallization of amorphous ice.

\item Low level activity due to equilibrium sublimation of exposed supervolatile ices cannot be excluded by our data.  However, the fraction of the surface actively sublimating in equilibrium with sunlight must be $f_A \lesssim 10^{-4}$, reflecting the instability of supervolatiles even in the 19 AU to 24 AU range.

\item Our search yielded no binaries.  Based on our sample and Poisson counting statistics, we conclude that the binary fraction is $<8$\% (3$\sigma$).  This is consistent with an origin of the HPOs in the dynamically hot population of the Kuiper belt and binary disruption caused by interaction with the giant planets \citep{2014MNRAS.437.2297B}.

\end{enumerate}

\acknowledgments
We thank the anonymous referee for the comments, specially on the phase coefficients. We are prompted to read the relevant papers which improve our understanding on the phase coefficients of centaurs. We thank Dr. Yoonyoung Kim for her discussions with us about the paper. Based on observations made under GO 15344  with the NASA/ESA Hubble Space Telescope, obtained at the Space Telescope Science Institute,  operated by the Association of Universities for Research in Astronomy, Inc., under NASA contract NAS 5-26555.

\clearpage

\startlongtable
\begin{deluxetable}{rlcccr}
\tablecolumns{1}
\tablewidth{0pc} 
\tablecaption{Orbital Parameters Sorted by Perihelion Distance\label{tobj}} 
\tablehead{ 
\multicolumn{2}{c}{Object} &
\colhead{q\tablenotemark{1}} &
\colhead{a\tablenotemark{2}} &
\colhead{e\tablenotemark{3}} &
\colhead{i\tablenotemark{4}} \\
\colhead{\#} &
\colhead{Name} &
\colhead{[AU]} &
\colhead{[AU]} &
\colhead{} &
\colhead{} 
}
%\decimalcolnumbers
\startdata 
    & (2014 JG80) &       15.175 &       21.286 &      0.287 &       32.8\\
471272  & (2011 FY9) &       15.184 &       59.366 &      0.744 &       37.8\\
  87555  & (2000 QB243) &       15.209 &       34.667 &      0.561 &       6.8\\
341275  & (2007 RG283) &       15.292 &       20.007 &      0.236 &       28.8\\
 & (2014 FB72) &       15.557 &       23.865 &      0.348 &       17.0\\
523753  & (2014 WV508) &       15.597 &       55.330 &      0.718 &       21.2\\
 & (2013 UR15) &       15.686 &       56.019 &      0.720 &       22.3\\
 & (2010 LO33) &       15.686 &       22.852 &      0.314 &       17.9\\
  & (2013 PU74) &       15.796 &       33.626 &      0.530 &       12.7\\
523746  & (2014 UT114) &       15.886 &       30.385 &      0.477 &       15.2\\
523720  & (2014 LN28) &       16.264 &       35.988 &      0.548 &       8.7\\
 514312  & (2016 AE193) &       16.522 &       31.356 &      0.473 &       10.2\\
 523673  & (2013 MZ11) &       16.765 &       24.169 &      0.306 &       6.4\\
523719 & (2014 LM28) &       16.771 &       262.038 &      0.936 &       84.8\\
 & (2015 BF515) &       16.970 &       20.423 &      0.169 &       28.3\\
 & (2014 GQ53) &       17.377 &       25.317 &      0.314 &       22.8\\
42355 Typhon  & (2002 CR46) &       17.512 &       37.784 &      0.537 &       2.4\\
44594 & (1999 OX3) &       17.572 &       32.263 &      0.455 &       2.6\\
 & (2007 BP102) &       17.722 &       23.938 &      0.260 &       64.8\\
 & (2014 JE80) &       17.934 &       90.071 &      0.801 &       28.2\\
  & (2014 XQ40) &       18.000 &       68.667 &      0.738 &       14.7\\
471149 & (2010 FB49) &       18.192 &       22.546 &      0.193 &       24.4\\
   & (2012 GU11) &       18.203 &       182.852 &      0.900 &       10.7\\
523686 & (2014 DB143) &       18.293 &       20.148 &     0.092 &       21.3\\
  & (2014 NX65) &       18.394 &       22.828 &      0.194 &       11.4\\
 & (2013 CE223) &       18.513 &       21.898 &      0.155 &       5.2\\
 & (2005 UN524) &       18.740 &       21.640 &      0.134 &       17.8\\
 & (2010 WG9) &       18.759 &       53.833 &      0.652 &       70.2\\
  523710  & (2014 JF80) &       18.764 &       33.656 &      0.442 &       13.8\\
523709  & (2014 JD80) &       18.982 &       25.239 &      0.248 &       39.1\\
471513  & (2012 CE17) &       18.985 &       21.578 &      0.120 &       5.9\\
 & (2013 RG98) &       19.284 &       23.243 &      0.170 &       46.0\\
 & (2008 AU138) &       20.128 &       32.323 &      0.377 &       42.8\\
463368  & (2012 VU85) &       20.209 &       29.407 &      0.313 &       15.0\\
 & (2013 FN28) &       20.331 &       35.542 &      0.428 &       8.6\\
   & (2011 FX62) &       20.367 &       48.256 &      0.578 &       18.2\\
 & (2006 UX184) &       20.423 &       38.195 &      0.465 &       37.4\\
33128  & (1998 BU48) &       20.523 &       33.382 &      0.385 &       14.2\\
 & (2004 VM131) &       20.630 &       67.839 &      0.696 &       14.1\\
 & (2002 PQ152) &       20.894 &       25.754 &      0.189 &       9.3\\
 & (2013 UE15) &       20.907 &       60.806 &      0.656 &       6.7\\
127546  & (2002 XU93) &       20.981 &       67.390 &      0.689 &       77.9\\
501214  & (2013 TC146) &       21.006 &       25.164 &      0.165 &       14.2\\
 316179  & (2010 EN65) &       21.066 &       30.785 &      0.316 &       19.2\\
 & (2008 KV42) &       21.171 &       41.858 &      0.494 &       103.5\\
 & (2014 SB349) &       21.220 &       25.845 &      0.179 &       15.5\\
  & (2014 GP53) &       21.290 &       26.772 &      0.205 &       14.3\\
  & (2013 MY11) &       21.511 &       51.646 &      0.583 &       14.5\\
 & (2010 TV191) &       21.694 &       24.856 &      0.127 &       11.5\\
  471155  & (2010 GF65) &       22.096 &       33.266 &      0.336 &       12.4\\
 & (2004 MW8) &       22.414 &       33.524 &      0.331 &       8.2\\
 & (2014 GR53) &       22.634 &       212.236 &      0.893 &       42.1\\
160427  & (2005 RL43) &       23.546 &       24.614 &     0.043 &       12.3\\
\enddata  
\tablenotetext{1}{ Perihelion distance}
\tablenotetext{2}{ Semi-major axis}
\tablenotetext{3}{ Orbital eccentricity}
\tablenotetext{4}{ Orbital inclination in degrees}
\end{deluxetable} 

\clearpage
\startlongtable
\begin{deluxetable}{rlccccc}
\tablecolumns{1}
\tablewidth{0pc} 
\tablecaption{Centaur Observations in Chronological Order\label{tobs}} 
\tablehead{ 
\multicolumn{2}{c}{Object} &
\colhead{Date [UT]\tablenotemark{1}} &
\colhead{$r_H$\tablenotemark{2}} &
\colhead{$\Delta$\tablenotemark{3}} &
\colhead{$\alpha$\tablenotemark{4}} &
\colhead{$V$\tablenotemark{5}} 
\\
\colhead{\#} &
\colhead{Name} &
\colhead{} &
\colhead{[AU]} &
\colhead{[AU]} &
\colhead{} &
\colhead{} 
}
\startdata 
 & (2006 UX184) & 2017-10-20 &       20.445 &       19.651 &       1.7 &       22.07\\
523753  & (2014 WV508) & 2017-10-20 &       15.879 &       16.083 &       3.5 &       21.91\\
 & (2004 MW8) & 2017-10-21 &       26.345 &       26.045 &       2.1 &       22.92\\
 & (2008 AU138) & 2017-10-30 &       44.079 &       44.149 &       1.3 &       23.44\\
 & (2013 UE15) & 2017-11-01 &       21.959 &       21.131 &       1.5 &       22.34\\
523746  & (2014 UT114) & 2017-11-11 &       16.182 &       15.247 &       1.2 &       20.84\\
341275  & (2007 RG283) & 2017-11-16 &       15.769 &       14.924 &       1.9 &       20.69\\
 & (2014 SB349) & 2017-12-03 &       23.778 &       22.937 &       1.3 &       23.35\\
471513  & (2012 CE17) & 2017-12-07 &       18.976 &       19.127 &       2.9 &       22.23\\
 & (2013 UR15) & 2017-12-12 &       17.978 &       17.230 &       2.1 &       23.71\\
501214  & (2013 TC146) & 2017-12-13 &       26.600 &       27.127 &       1.8 &       21.06\\
 & (2010 TV191) & 2017-12-15 &       27.989 &       27.056 &      0.6 &       22.81\\
 & (2004 VM131) & 2017-12-16 &       31.421 &       30.448 &      0.3 &       23.21\\
 & (2015 BF515) & 2017-12-17 &       17.745 &       16.935 &       1.9 &       22.41\\
 & (2010 WG9) & 2017-12-19 &       23.056 &       22.083 &      0.4 &       21.71\\
 & (2013 RG98) & 2017-12-20 &       21.842 &       21.269 &       2.1 &       22.75\\
 & (2005 UN524) & 2017-12-31 &       18.744 &       17.798 &      0.8 &       22.06\\
 & (2007 BP102) & 2018-01-13 &       18.732 &       19.031 &       2.8 &       23.55\\
523719 & (2014 LM28) & 2018-02-01 &       16.773 &       17.161 &       3.1 &       22.47\\
 & (2014 JE80) & 2018-02-13 &       17.978 &       18.413 &       2.8 &       22.81\\
 & (2014 GR53) & 2018-02-16 &       22.634 &       22.565 &       2.5 &       21.90\\
 & (2002 PQ152) & 2018-02-18 &       23.652 &       24.088 &       2.1 &       23.49\\
 & (2008 KV42) & 2018-02-25 &       26.604 &       26.523 &       2.1 &       23.26\\
 & (2013 FN28) & 2018-03-01 &       20.331 &       19.508 &       1.6 &       21.83\\
471272  & (2011 FY9) & 2018-03-04 &       19.879 &       19.312 &       2.4 &       22.13\\
 & (2014 FB72) & 2018-03-18 &       17.887 &       17.294 &       2.6 &       21.21\\
463368  & (2012 VU85) & 2018-03-21 &       24.413 &       24.572 &       2.3 &       22.89\\
471149 & (2010 FB49) & 2018-04-11 &       26.435 &       25.440 &      0.3 &       21.53\\
523709  & (2014 JD80) & 2018-04-12 &       19.692 &       20.023 &       2.7 &       22.45\\
127546  & (2002 XU93) & 2018-04-14 &       23.564 &       23.539 &       2.4 &       21.94\\
42355 Typhon  & (2002 CR46) & 2018-04-15 &       21.606 &       20.615 &      0.4 &       21.14\\
 & (2010 LO33) & 2018-04-19 &       16.996 &       16.610 &       3.2 &       21.60\\
523686 & (2014 DB143) & 2018-04-30 &       19.385 &       18.580 &       1.8 &       21.96\\
 & (2014 GQ53) & 2018-05-07 &       19.055 &       18.149 &       1.4 &       22.29\\
44594 & (1999 OX3) & 2018-05-25 &       18.497 &       19.123 &       2.4 &       20.19\\
523720  & (2014 LN28) & 2018-09-13 &       17.254 &       16.363 &       1.6 &       20.78\\
33128  & (1998 BU48) & 2019-03-08 &       35.370 &       34.402 &      0.4 &       22.91\\
 & (2013 CE223) & 2019-03-09 &       23.847 &       22.859 &      0.3 &       22.60\\
160427  & (2005 RL43) & 2019-03-10 &       24.404 &       24.746 &       2.2 &       21.89\\
  & (2014 XQ40) & 2019-03-21 &       18.751 &       17.820 &       1.1 &       21.74\\
 316179  & (2010 EN65) & 2019-03-29 &       26.955 &       25.990 &      0.6 &       21.45\\
  & (2013 MY11) & 2019-04-23 &       23.161 &       23.632 &       2.2 &       22.45\\
 523673  & (2013 MZ11) & 2019-04-24 &       19.988 &       20.435 &       2.6 &       21.50\\
  & (2014 GP53) & 2019-05-03 &       21.860 &       20.889 &      0.7 &       22.40\\
   & (2012 GU11) & 2019-05-06 &       24.631 &       23.658 &      0.6 &       22.99\\
 514312  & (2016 AE193) & 2019-05-20 &       17.359 &       17.630 &       3.2 &       21.14\\
   & (2014 JG80) & 2019-05-31 &       27.338 &       26.361 &      0.6 &       22.16\\
  & (2013 PU74) & 2019-05-31 &       16.416 &       16.798 &       3.2 &       22.72\\
  87555  & (2000 QB243) & 2019-06-02 &       31.155 &       31.630 &       1.6 &       23.81\\
  471155  & (2010 GF65) & 2019-07-02 &       23.192 &       22.347 &       1.4 &       21.04\\
   & (2011 FX62) & 2019-08-28 &       25.101 &       25.476 &       2.1 &       21.29\\
  523710  & (2014 JF80) & 2019-10-03 &       18.776 &       17.890 &       1.4 &       21.69\\
  & (2014 NX65) & 2019-10-06 &       18.931 &       17.971 &      0.9 &       22.43\\
\enddata  
\tablenotetext{1}{ Observing date}
\tablenotetext{2}{ Heliocentric distance}
\tablenotetext{3}{ Geocentric distance}
\tablenotetext{4}{ Phase angle}
\tablenotetext{5}{ Apparent magnitude from 0.2\arcsec~aperture}
\end{deluxetable} 

\clearpage
\startlongtable
\begin{deluxetable}{rlcccrrrr}
\tablecolumns{1}
\tablewidth{0pc} 
\tablecaption{Derived Quantities in Order of HPO Designated Names\label{tquantities}} 
\tablehead{ 
\multicolumn{2}{c}{Object} &
\colhead{Date [UT]\tablenotemark{1}} &
\colhead{$H_V$\tablenotemark{2}} &
\colhead{$r_e$\tablenotemark{3}} &
\colhead{C\tablenotemark{4}} &
\colhead{$M\tablenotemark{5}$} &
\colhead{$\tau\tablenotemark{6}$} &
\colhead{$\dot{M}$\tablenotemark{7}} \\
\colhead{} &
\colhead{} &
\colhead{} &
\colhead{} &
\colhead{[km]} &
\colhead{[m$^2$]} &
\colhead{[kg]} &
\colhead{[hr]} &
\colhead{[kg s$^{-1}$]}}
\startdata 
 33128  & (1998 BU48) & 2019-03-08 &       7.5 &       69 & 1.5$\times 10^{8}$ & 2.0$\times 10^{6}$ &       23.1 &       24\\
44594 & (1999 OX3) & 2018-05-25 &       7.3 &       74 & 4.3$\times 10^{7}$ & 5.7$\times 10^{5}$ &       12.8 &       12\\
  87555  & (2000 QB243) & 2019-06-02 &       8.7 &       38 & 1.1$\times 10^{8}$ & 1.5$\times 10^{6}$ &       21.2 &       20\\
42355   & (2002 CR46) & 2018-04-15 &       7.9 &       57 & 1.3$\times 10^{8}$ & 1.7$\times 10^{6}$ &       13.8 &       34\\
127546  & (2002 XU93) & 2018-04-14 &       8.1 &       52 & 2.1$\times 10^{8}$ & 2.7$\times 10^{6}$ &       15.8 &       48\\
 & (2002 PQ152) & 2018-02-18 &       9.6 &       26 & 3.1$\times 10^{7}$ & 4.2$\times 10^{5}$ &       16.2 &       7\\
 & (2004 MW8) & 2017-10-21 &       8.6 &       40 & 3.3$\times 10^{7}$ & 4.4$\times 10^{5}$ &       17.5 &       7\\
 & (2004 VM131) & 2017-12-16 &       8.3 &       47 & 1.5$\times 10^{8}$ & 2.1$\times 10^{6}$ &       20.4 &       28\\
 & (2005 UN524) & 2017-12-31 &       9.4 &       28 & 1.7$\times 10^{7}$ & 2.3$\times 10^{5}$ &       11.9 &       5\\
160427  & (2005 RL43) & 2019-03-10 &       7.9 &       57 & 8.7$\times 10^{7}$ & 1.2$\times 10^{6}$ &       16.6 &       19\\
 & (2006 UX184) & 2017-10-20 &       8.9 &       35 & 8.7$\times 10^{7}$ & 1.2$\times 10^{6}$ &       13.2 &       24\\
 & (2007 BP102) & 2018-01-13 &       10.6 &       16 & 1.9$\times 10^{7}$ & 2.5$\times 10^{5}$ &       12.8 &       5\\
341275  & (2007 RG283) & 2017-11-16 &       8.7 &       38 & 2.1$\times 10^{7}$ & 2.8$\times 10^{5}$ &       10.0 &       8\\
 & (2008 KV42) & 2018-02-25 &       8.9 &       36 & 2.4$\times 10^{7}$ & 3.2$\times 10^{5}$ &       17.8 &       5\\
 & (2008 AU138) & 2017-10-30 &       6.9 &       88 & 1.3$\times 10^{8}$ & 1.8$\times 10^{6}$ &       29.6 &       17\\
  471155  & (2010 GF65) & 2019-07-02 &       7.4 &       71 & 7.1$\times 10^{7}$ & 9.5$\times 10^{5}$ &       15.0 &       18\\
 & (2010 WG9) & 2017-12-19 &       8.2 &       50 & 7.8$\times 10^{6}$ & 1.0$\times 10^{5}$ &       14.8 &       2\\
 316179  & (2010 EN65) & 2019-03-29 &       7.2 &       78 & 2.1$\times 10^{8}$ & 2.8$\times 10^{6}$ &       17.4 &       44\\
471149 & (2010 FB49) & 2018-04-11 &       7.4 &       71 & 5.4$\times 10^{3}$ & 7.2$\times 10^{1}$ &       17.1 &      1\tablenotemark{8}\\
 & (2010 TV191) & 2017-12-15 &       8.4 &       45 & 9.2$\times 10^{7}$ & 1.2$\times 10^{6}$ &       18.2 &       19\\
 & (2010 LO33) & 2018-04-19 &       9.1 &       31 & 10.$\times 10^{7}$ & 1.3$\times 10^{6}$ &       11.2 &       33\\
471272  & (2011 FY9) & 2018-03-04 &       9.1 &       33 & 7.7$\times 10^{7}$ & 1.0$\times 10^{6}$ &       13.0 &       22\\
   & (2011 FX62) & 2019-08-28 &       7.1 &       80 & 9.0$\times 10^{7}$ & 1.2$\times 10^{6}$ &       17.1 &       19\\
463368  & (2012 VU85) & 2018-03-21 &       8.9 &       36 & 2.2$\times 10^{7}$ & 3.0$\times 10^{5}$ &       16.5 &       5\\
   & (2012 GU11) & 2019-05-06 &       9.1 &       32 & 9.0$\times 10^{7}$ & 1.2$\times 10^{6}$ &       15.9 &       21\\
471513  & (2012 CE17) & 2017-12-07 &       9.3 &       30 & 1.8$\times 10^{7}$ & 2.5$\times 10^{5}$ &       12.8 &       5\\
  & (2013 MY11) & 2019-04-23 &       8.6 &       40 & 2.8$\times 10^{7}$ & 3.7$\times 10^{5}$ &       15.9 &       6\\
 & (2013 UE15) & 2017-11-01 &       8.9 &       35 & 2.9$\times 10^{7}$ & 3.8$\times 10^{5}$ &       14.2 &       8\\
 & (2013 RG98) & 2017-12-20 &       9.3 &       30 & 3.0$\times 10^{7}$ & 4.0$\times 10^{5}$ &       14.3 &       8\\
 & (2013 UR15) & 2017-12-12 &       11.1 &       13 & 1.5$\times 10^{7}$ & 2.0$\times 10^{5}$ &       11.6 &       5\\
501214  & (2013 TC146) & 2017-12-13 &       6.7 &       99 & 4.2$\times 10^{8}$ & 5.6$\times 10^{6}$ &       18.2 &       85\\
 & (2013 FN28) & 2018-03-01 &       8.7 &       38 & 2.5$\times 10^{7}$ & 3.3$\times 10^{5}$ &       13.1 &       7\\
  & (2013 PU74) & 2019-05-31 &       10.3 &       18 & 1.4$\times 10^{7}$ & 1.8$\times 10^{5}$ &       11.3 &       5\\
 523673  & (2013 MZ11) & 2019-04-24 &       8.3 &       47 & 8.6$\times 10^{8}$ & 1.1$\times 10^{7}$ &       13.7 &       231\\
 & (2013 CE223) & 2019-03-09 &       8.9 &       35 & 7.0$\times 10^{7}$ & 9.4$\times 10^{5}$ &       15.3 &       17\\
  & (2014 NX65) & 2019-10-06 &       9.7 &       24 & 3.5$\times 10^{7}$ & 4.7$\times 10^{5}$ &       12.1 &       11\\
523753  & (2014 WV508) & 2017-10-20 &       9.7 &       25 & 5.8$\times 10^{6}$ & 7.8$\times 10^{4}$ &       10.8 &       2\\
523746  & (2014 UT114) & 2017-11-11 &       8.8 &       37 & 6.4$\times 10^{6}$ & 8.5$\times 10^{4}$ &       10.2 &       2\\
 & (2014 SB349) & 2017-12-03 &       9.6 &       26 & 7.1$\times 10^{7}$ & 9.5$\times 10^{5}$ &       15.4 &       17\\
523719 & (2014 LM28) & 2018-02-01 &       10.0 &       21 & 2.0$\times 10^{7}$ & 2.7$\times 10^{5}$ &       11.5 &       6\\
 & (2014 JE80) & 2018-02-13 &       10.0 &       21 & 4.2$\times 10^{7}$ & 5.7$\times 10^{5}$ &       12.4 &       13\\
 & (2014 GR53) & 2018-02-16 &       8.2 &       49 & 4.4$\times 10^{7}$ & 5.9$\times 10^{5}$ &       15.1 &       11\\
 & (2014 FB72) & 2018-03-18 &       8.6 &       41 & 3.4$\times 10^{7}$ & 4.5$\times 10^{5}$ &       11.6 &       11\\
523709  & (2014 JD80) & 2018-04-12 &       9.3 &       29 & 4.5$\times 10^{7}$ & 6.0$\times 10^{5}$ &       13.4 &       12\\
523686 & (2014 DB143) & 2018-04-30 &       9.1 &       33 & 7.1$\times 10^{7}$ & 9.4$\times 10^{5}$ &       12.5 &       21\\
 & (2014 GQ53) & 2018-05-07 &       9.5 &       27 & 2.0$\times 10^{8}$ & 2.7$\times 10^{6}$ &       12.2 &       61\\
523720  & (2014 LN28) & 2018-09-13 &       8.4 &       44 & 2.7$\times 10^{7}$ & 3.6$\times 10^{5}$ &       11.0 &       9\\
  & (2014 XQ40) & 2019-03-21 &       9.0 &       33 & 5.1$\times 10^{6}$ & 6.8$\times 10^{4}$ &       12.0 &       2\\
  & (2014 GP53) & 2019-05-03 &       9.1 &       33 & 7.1$\times 10^{7}$ & 9.5$\times 10^{5}$ &       14.0 &       19\\
   & (2014 JG80) & 2019-05-31 &       7.8 &       58 & 2.6$\times 10^{7}$ & 3.5$\times 10^{5}$ &       17.7 &       5\\
  523710  & (2014 JF80) & 2019-10-03 &       9.0 &       34 & 1.2$\times 10^{3}$ & 1.6$\times 10^{1}$ &       12.0 &      1\tablenotemark{8}\\
 & (2015 BF515) & 2017-12-17 &       9.9 &       22 & 2.7$\times 10^{7}$ & 3.6$\times 10^{5}$ &       11.4 &       9\\
 514312  & (2016 AE193) & 2019-05-20 &       8.5 &       42 & 3.3$\times 10^{7}$ & 4.5$\times 10^{5}$ &       11.8 &       10\\
\enddata  
\tablenotetext{1}{ Observing date}
\tablenotetext{2}{ Absolute magnitude, computed from Equation (\ref{H})}
\tablenotetext{3}{ Nuclear radius, computed from Equation (\ref{r_e})}
\tablenotetext{4}{ Cross-section of dust projected within the aperture}
\tablenotetext{5}{ Dust mass within aperture}
\tablenotetext{6}{ Aperture residence time}
\tablenotetext{7}{ Upper limit mass loss from Equation (\ref{dmbdt})}
\tablenotetext{8}{Mass loss rate are rounded to 1}
\end{deluxetable} 

\clearpage
\begin{deluxetable}{cclccc} 
\tablecolumns{1}
\tablewidth{0pc} 
\tablecaption{Statistics of Coma Measurements\label{tmagdiff}} clearpage
\tablehead{ 
\colhead{$(i,j)$\tablenotemark{a}} &
\colhead{Radii} &
\colhead{} &
\colhead{$(V_i-V_j)_{HPO}$} &
\colhead{$(V_i-V_j)_\star$} &
\colhead{$\Delta V_{i,j}$\tablenotemark{b}}
}
\startdata 
0,1 & 0.2\arcsec~- 0.4\arcsec~ & mean & 0.095$\pm$0.004 &0.085$\pm$0.003 & 0.010$\pm$0.005\\
  & &median & 0.088 & 0.082 & 0.006\\
% & $\sigma$\tablenotemark{b} & 0.038& 0.016&0.039 \\
0,2 & 0.2\arcsec~- 0.8\arcsec~ & mean & 0.145$\pm$0.005& 0.136$\pm$0.005 &0.009$\pm$0.007\\
   &&median & 0.138& 0.127& 0.011 \\
%      & $\sigma$ & 0.066& 0.025 & 0.066
1,2 & 0.4\arcsec~- 0.8\arcsec~ & mean & 0.050$\pm$0.004 & 0.051$\pm$0.004 &-0.001$\pm$0.006\\
   & &median & 0.046 & 0.048  & -0.002\\
%   & $\sigma$& 0.040&0.020 &0.040\\
%I THINK THE SIGMAS ON THE MAG DIFFERENCE ARE WRONG (LAST COLUMN)
\enddata  
\tablenotetext{a}{Annuli }
%\tablenotetext{b}{Standard deviation}
\tablenotetext{b}{Excess magnitude from Equation (\ref{vc})}
\end{deluxetable} 

\clearpage

\clearpage 

\clearpage

\clearpage

%\clearpage
%
%
%\begin{figure}
%\epsscale{1.0}
%\plotone{Apparent-Geor.pdf}
%\caption{Apparent magnitudes of Centaurs as function of geocentric distance. The furthest object is 2008 AU138. 
%\label{apparent-geor}}
%\end{figure}

\clearpage

\clearpage

%\clearpage
%
%
%\begin{figure}
%\epsscale{1.0}
%\plotone{compare_absmag.pdf}
%\caption{Absolute magnitudes of Centaurs listed in Horizons as function of the absolute magnitude obtained from this work.
%\label{compare-absmag}}
%\end{figure}
% DO WE NEED TO SHOW BOTH APP MAG AND ABS MAGS?  ARE THEY NOT THE SAME THING?

\clearpage

\clearpage

\clearpage

\clearpage

\clearpage

\end{CJK*}
\end{document}